\documentclass[11pt]{article}
\usepackage[preprint]{acl}
\usepackage{times,latexsym}
\usepackage[T1]{fontenc}
\usepackage[utf8]{inputenc}
\usepackage{microtype,inconsolata}
\usepackage{graphicx,amsmath,amssymb,booktabs,tabularx,multirow}
\usepackage{placeins}
\usepackage{tikz}
\usetikzlibrary{arrows.meta,positioning,fit,calc}
\title{TTS-Q: Speaker-Aware Post-Training Quantization\\of Text-to-Speech Diffusion Transformers}
\author{%
  \textbf{Xiaoqin Feng\textsuperscript{1,*},
  Yuqi Li\textsuperscript{1,2,*},
  Weilun Feng\textsuperscript{3,$\dagger$},
  Yi-Cheng Lin\textsuperscript{4,$\dagger$},}\\
  \textbf{Kai Li\textsuperscript{5},
  Xianglong Wang\textsuperscript{1},
  Kuo Yang\textsuperscript{1},
  Chuanguang Yang\textsuperscript{3},}\\
  \textbf{Yingli Tian\textsuperscript{2},
  Wei Li\textsuperscript{1,$\ddagger$},
  Zhongwei Cheng\textsuperscript{1,$\ddagger$},
  Lin Chen\textsuperscript{1,$\ddagger$}}\\[0.5em]
  {\normalfont\small\textsuperscript{1}Wyze Inc., USA}\\
  {\normalfont\small\textsuperscript{2}The City College of New York, City University of New York, USA}\\
  {\normalfont\small\textsuperscript{3}Institute of Computing Technology, Chinese Academy of Sciences, China}\\
  {\normalfont\small\textsuperscript{4}National Taiwan University, Taiwan\quad
  \textsuperscript{5}Tsinghua University, China}\\[0.4em]
  {\normalfont\small\textsuperscript{*}Equal contribution.\quad
  \textsuperscript{$\dagger$}Corresponding authors.\quad
  \textsuperscript{$\ddagger$}Project leaders.}
}
\begin{document}
\maketitle
\begin{abstract}
Zero-shot text-to-speech (TTS) generates speech in a target voice from a short reference recording. The large size of speech diffusion transformers motivates post-training quantization (PTQ), which must preserve both intelligibility and speaker identity. This raises two questions for calibration: which reference voices to represent and how to preserve voice identity in the generated speech. We propose TTS-Q, a speaker-aware PTQ method that combines coverage-guided calibration sampling with speaker identity distillation. Coverage in speaker embedding space guides the selection of reference conditions for reconstruction. Subsequent calibration uses the full-precision teacher's synthesized speech to supervise voice identity. Experiments on LongCat-AudioDiT and F5-TTS at W4A4 show that TTS-Q consistently improves speaker similarity and predicted naturalness over reconstruction with randomly selected conditions, while generally reducing text error. It also outperforms the evaluated external baselines in both quality metrics. In deployment tests, LongCat runs 1.66--2.34$\times$ faster than its native FP32 reference during warm inference. F5 remains slower than its native FP16 reference.
\end{abstract}
\section{Introduction}
Zero-shot text-to-speech (TTS) generates new utterances in a voice specified by a short reference recording. Neural codec language models \citep{valle} and diffusion or flow-matching models \citep{naturalspeech2,voicebox,e2tts} have advanced this capability. Recent diffusion transformers (DiTs) generate speech in mel-spectrogram space \citep{f5tts} or waveform latent space \citep{longcat}, but their model size motivates compression. Post-training quantization (PTQ) reduces weight and activation precision using a small calibration set. For TTS, the compressed model must preserve both the spoken content and the voice specified by the reference recording.

\begin{figure*}[t]
\centering
\includegraphics[width=\textwidth]{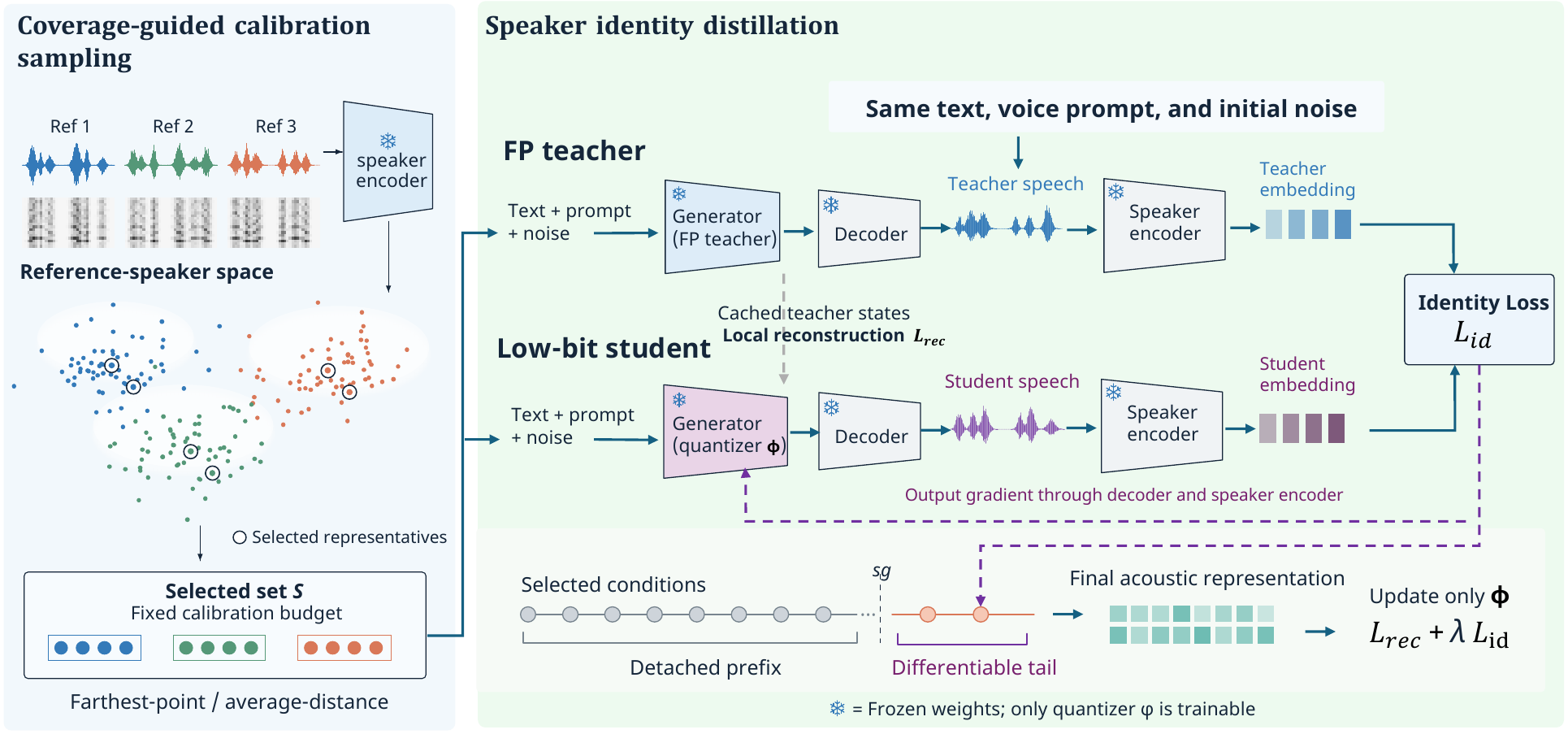}
\caption{Overview of TTS-Q. Coverage-guided calibration sampling selects conditions in a reference-speaker embedding space. After initial local reconstruction, speaker identity distillation compares teacher and student speech under matched text, voice prompt, and initial noise. Local reconstruction remains active on cached teacher states. The dashed purple path carries identity gradients through the frozen decoder and speaker encoder. The lower inset shows a detached prefix (student for LongCat, teacher for F5) and differentiable sampling tail. Only quantization parameters are updated. Waveforms and embedding geometry are schematic.}
\label{fig:overview}
\end{figure*}

PTQ methods limit numerical error through local reconstruction or learned transformations \citep{adaround,brecq,flatquant}. Diffusion-specific methods also account for changes across sampling steps and feature channels \citep{qdiffusion,ptq4dit,qdit}. For zero-shot TTS, constructing a calibration set involves choosing both sampling states and the reference voices that condition them. Sampling more states from the same recordings broadens temporal coverage without changing the voices represented. Random selection does not explicitly control voice coverage under a limited budget. Calibration data are known to affect compression quality \citep{calibration_data}; here, we investigate coverage in speaker embedding space as a criterion for choosing reference conditions.

The calibration objective determines what the model preserves on these conditions. Local reconstruction matches intermediate computations, but does not directly constrain voice identity after sampling and decoding. A loss on speaker embeddings of the generated speech can provide this supervision. Since quantization aims to retain the pretrained model's behavior, we use speech generated by the full-precision teacher as the identity target for each condition.

We introduce TTS-Q, a speaker-aware PTQ method for speech DiTs with two calibration stages (Figure~\ref{fig:overview}). First, we select reference conditions by their coverage in a frozen speaker embedding space and calibrate the model through local reconstruction. We then continue calibration with an identity loss between teacher and student speech, while retaining the reconstruction loss. Only quantization parameters are optimized; the pretrained generator remains fixed. Our contributions are:
\begingroup
\setlength{\parskip}{0pt}
\newcommand{\introcontribution}[1]{\par\vspace{1.5pt}\noindent\hangindent=1.25em\hangafter=1\makebox[1.25em][l]{$\bullet$}#1}
\introcontribution{\textbf{Coverage-Guided Calibration Sampling} selects calibration conditions under a fixed budget by their coverage in speaker embedding space.}
\introcontribution{\textbf{Speaker Identity Distillation} uses the teacher's synthesized speech to supervise voice identity during further calibration, complementing local reconstruction.}
\introcontribution{Across three models from two speech DiT families at W4A4, TTS-Q consistently improves speaker similarity and predicted naturalness over reconstruction with randomly selected conditions, while generally reducing text error. Deployment evaluation further demonstrates storage savings across all models and warm-inference acceleration on LongCat over its native FP32 reference.}
\par\endgroup

\section{Related Work}
\label{sec:related}
\newcommand{\rwheading}[1]{\par\addvspace{0.4ex}\noindent\textbf{#1}\hspace{0.5em}}
\rwheading{Speech synthesis and compression.}
Zero-shot speech diffusion systems generate different acoustic representations: F5-TTS produces mel-spectrograms, whereas LongCat-AudioDiT operates in a waveform latent space \citep{f5tts,longcat}. TTS compression has also been studied through quantization-aware training of compact architectures, including ternary-weight BitTTS \citep{bittts}. We study post-training W4A4 calibration of pretrained speech DiTs, with fixed generator weights and each model's original acoustic representation and decoder.

\rwheading{Post-training quantization for iterative generators.}
General PTQ methods reduce numerical distortion through second-order compensation, activation-aware scaling, rotations, and learned transformations \citep{gptq,awq,quarot,flatquant}. Diffusion PTQ additionally accounts for changes along the sampling trajectory. Earlier work calibrates across timesteps or reconstructs diffusion-specific blocks, while DiT methods address channel imbalance, quantization granularity, and dynamic activations \citep{qdiffusion,ptq4dit,qdit,viditq}. Work on audio DiTs studies timestep-aware smoothing and low-rank error compensation \citep{audio_ptq}. We use FlatQuant's learned transforms as the reconstruction backend and study the choice of calibration conditions and supervision.

\rwheading{Calibration data for audio generation.}
The sensitivity of compression to calibration data has motivated analyses of sampled data and model-generated self-calibration \citep{calibration_data,self_calibration}. Core-set methods select subsets that represent a candidate pool under a fixed budget \citep{coreset}. For audio diffusion, PTQ4ADM augments prompts to cover modalities and attributes and constructs activation-aware calibration sets \citep{ptq4adm}. S$^2$Q-VDiT selects video calibration samples by combining denoising informativeness with a Hessian-based measure of quantization sensitivity \citep{s2qvdit}. TTS-Q selects reference conditions by their coverage in speaker embedding space.

\rwheading{Distillation and speaker identity.}
Q-VDiT preserves temporal relations during video quantization \citep{qvdit}. S$^2$Q-VDiT uses attention to weight token reconstruction losses \citep{s2qvdit}. TTS-Q instead compares speaker embeddings of the final synthesized speech, supervising voice identity beyond intermediate reconstruction. We build on speaker-consistency supervision in TTS \citep{yourtts}, using a pretrained speaker encoder \citep{ecapa}. The full-precision generator supplies the target voice for each utterance, while truncated backpropagation through sampling \citep{draft} allows the identity loss to guide quantization parameters.

\section{Method}
\label{sec:method}
TTS-Q selects calibration conditions by reference-speaker coverage and uses teacher-generated speech on those conditions to supervise output identity (Figure~\ref{fig:overview}). We quantize DiT linear layers to W4A4 using a FlatQuant-style learned-transform parameterization \citep{flatquant}, with per-output-channel weight scales and dynamic per-token activation scales. The trainable parameters $\phi$ comprise the input transforms, channel-balancing scales, and clipping parameters. Pretrained weights remain frozen, and operations outside the quantized linear layers retain their native precision. Quantizer details are given in Appendix~\ref{sec:implementation}.

Let $c$ contain the target text, a reference recording and its transcript, and initial generation noise. The full-precision teacher produces waveform $G(c)$. A frozen speaker encoder $E$ represents both reference recordings for selection and generated speech for distillation. The selected conditions are shared by local reconstruction and identity supervision.

\subsection{Coverage-Guided Calibration Sampling}
\label{sec:coverage}
With a limited calibration budget, we seek reference recordings that cover the candidate voices in speaker embedding space. Following core-set selection \citep{coreset}, we measure each candidate's distance to the selected set.

Let $\mathcal P$ index the candidate conditions, and let $a_i$ be the reference recording of condition $c_i$. Assuming nonzero speaker embeddings, we normalize each as $u_i=E(a_i)/\|E(a_i)\|_2$. For a nonempty selected index set $S\subseteq\mathcal P$, define the distance to the nearest selected condition as
\begin{equation}
 r_i(S)=\min_{j\in S}d_{ij},\qquad
 d_{ij}=\max\{0,1-u_i^\top u_j\}.
 \label{eq:uncovered}
\end{equation}
A large $r_i(S)$ indicates that the candidate's reference embedding is far from the selected set in speaker space.

For a budget $1\leq K\leq|\mathcal P|$, we seek to minimize one of two coverage costs:
\begin{equation}
 \begin{aligned}
 C_{\max}(S)&=\max_{i\in\mathcal P}r_i(S),\\
 C_{\mathrm{avg}}(S)&=\frac{1}{|\mathcal P|}\sum_{i\in\mathcal P}r_i(S),
 \end{aligned}\qquad |S|=K.
 \label{eq:coverage}
\end{equation}
The maximum cost emphasizes the least covered reference embedding, whereas the average cost gives more weight to densely populated regions of the pool. We choose between these geometric objectives for each model using development-set speech quality. Section~\ref{sec:selection-analysis} reports the resulting coverage and downstream quality.

For $C_{\max}$, farthest-point sampling adds the unselected candidate with the largest $r_j(S)$, starting from a condition determined by a fixed pseudorandom ordering.

\begin{figure}[t]
\centering
\includegraphics[width=\linewidth]{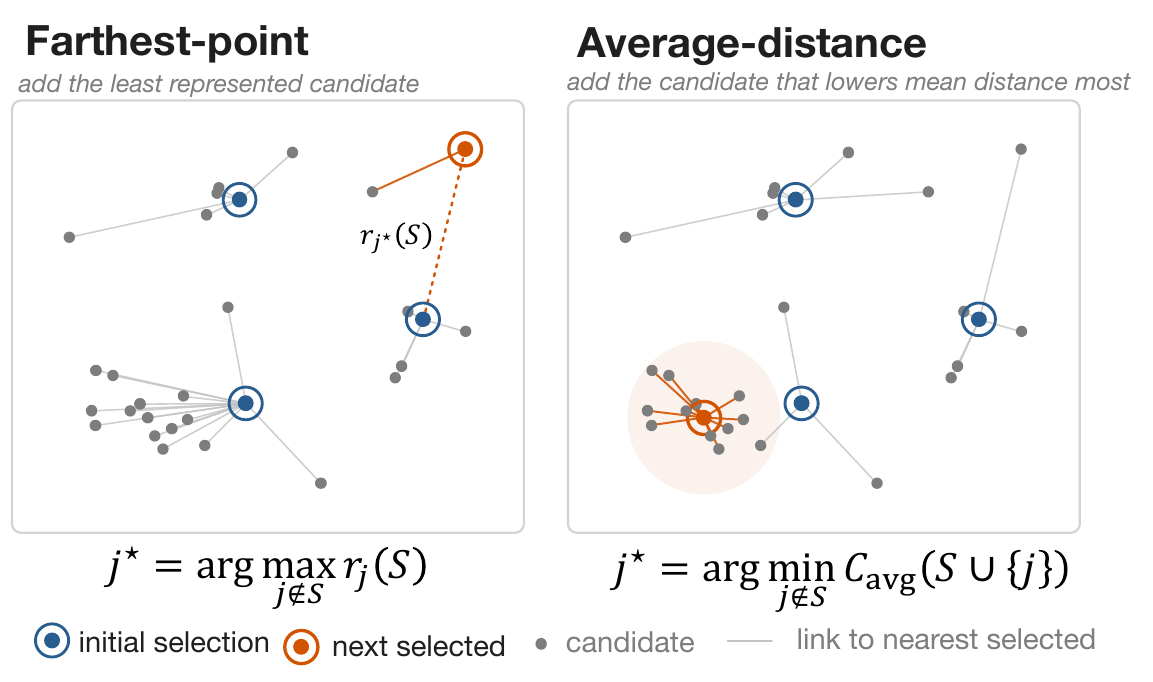}
\caption{One selection step in a schematic Euclidean space. Blue points form the current set $S$; orange marks the next candidate $j^\star$ selected from $\mathcal P\setminus S$. Solid links show nearest assignments after the addition; the dashed segment shows its distance to $S$ before the addition. The shared starting set is illustrative; initialization follows the rules in the text.}
\label{fig:coverage-geometry}
\end{figure}

For $C_{\mathrm{avg}}$, selection starts at the medoid, an index minimizing $|\mathcal P|^{-1}\sum_{i\in\mathcal P}d_{ij}$, and adds the candidate that minimizes $C_{\mathrm{avg}}(S\cup\{j\})$. Figure~\ref{fig:coverage-geometry} illustrates the two choices. Both rules use deterministic tie-breaking and stop at $|S|=K$.

Each selected recording retains its associated text and generation condition. Teacher states collected under these conditions provide reconstruction targets, and teacher speech provides the identity targets below. Candidate-pool construction and temporal sampling are specified in Appendix~\ref{sec:implementation}.

\paragraph{Initial reconstruction.}
For reconstruction unit $b$, let $h_b$ be its cached teacher input and $f_b$ and $f^q_{b,\phi}$ its original and quantized functions. Initial calibration minimizes
\begin{equation}
 \ell_b(\phi;h_b)=
 \frac{\|f^q_{b,\phi}(h_b)-f_b(h_b)\|^2_{\Omega}}
 {\max(\|f_b(h_b)\|^2_{\Omega},\epsilon)},
 \label{eq:rec-detail}
\end{equation}
where $\|\cdot\|^2_{\Omega}$ averages squared magnitude over valid tokens and channels, excluding padding, and $\epsilon=10^{-8}$. Units include transformer blocks and remaining DiT modules with linear layers. Each uses teacher inputs, so preceding quantization errors do not enter its local target. Summing $\ell_b$ over units at each teacher-state visit gives $\mathcal L_{\mathrm{rec}}$.

\subsection{Speaker Identity Distillation}
\label{sec:speaker}
We use speech generated by the full-precision teacher as the identity target to preserve its voice after quantization. The reference recording specifies the requested voice, and the teacher output captures how the full-precision model synthesizes it for the given utterance. Teacher and student use the same text, reference prompt, and initial noise, holding the generation condition fixed.

Let $\widetilde G_\phi(c)$ be the student waveform on the calibration sampling path described below. For nonzero embeddings, define cosine dissimilarity as $D(v,w)=1-v^\top w/(\|v\|_2\|w\|_2)$. The identity loss is
\begin{equation}
 \mathcal L_{\mathrm{id}}(c;\phi)=
 D\!\left(E(\widetilde G_\phi(c)),
          \operatorname{sg}[E(G(c))]\right),
 \label{eq:speaker}
\end{equation}
where $\operatorname{sg}$ denotes stop-gradient. Comparing speaker embeddings directs the loss toward voice identity without requiring sample-wise waveform agreement.

The second calibration stage starts from the checkpoint obtained by local reconstruction and updates the quantization parameters on the same conditions with the combined objective
\begin{equation}
 \mathcal L_{\mathrm{cal}}(c;\phi)=
 \mathcal L_{\mathrm{rec}}(c;\phi)
 +\lambda\mathcal L_{\mathrm{id}}(c;\phi),
 \label{eq:calibration}
\end{equation}
where $\lambda\geq0$ weights the identity term. Once per condition, we compute the identity gradient at a designated reconstruction visit, before updating any units at that visit. Each unit combines its local reconstruction gradient with the corresponding identity gradient weighted by $\lambda$, then clips the combined gradient and takes an optimizer step. Other visits use only reconstruction gradients.

We use truncated backpropagation through sampling \citep{draft}. Identity gradients pass through the frozen speaker encoder, waveform decoder, and final two sampling updates to $\phi$, with a straight-through estimator for quantization. LongCat starts these updates from a detached state on the current student trajectory. F5 instead uses cached teacher states to avoid recomputing the student prefix, so its calibration waveform does not reflect quantization errors accumulated in the earlier student sampling steps. The effect of this approximation on final quality has not been isolated. Appendix~\ref{sec:implementation} gives the sampling equations, injection schedule, and optimizer settings.

\section{Experiments}
\label{sec:experiments}
We compare TTS-Q with existing quantization methods, then examine the effects of condition selection and identity distillation. The evaluation covers speech quality, model storage, and inference latency.

\begin{table*}[!t]
\centering\footnotesize
\setlength{\tabcolsep}{0.8pt}
\begin{tabular}{l@{\hspace{3pt}}l*{9}{c}}
\toprule
&&\multicolumn{3}{c}{Seed English}&\multicolumn{3}{c}{Seed Mandarin}&\multicolumn{3}{c}{Seed Mandarin hard}\\
\cmidrule(lr){3-5}\cmidrule(lr){6-8}\cmidrule(lr){9-11}
Model & Method & WER$\downarrow$ & SIM$\uparrow$ & UTMOS$\uparrow$ & CER$\downarrow$ & SIM$\uparrow$ & UTMOS$\uparrow$ & CER$\downarrow$ & SIM$\uparrow$ & UTMOS$\uparrow$\\
\midrule
LongCat-1B & Full Prec & 2.403 & 0.7623 & 3.789 & 1.263 & 0.8118 & 3.161 & 7.058 & 0.7891 & 2.997 \\
\cmidrule(lr){2-11}
 & RTN & 125.333 & 0.0420 & 1.287 & 100.243 & 0.0145 & 1.310 & 100.692 & -0.0063 & 1.286 \\
 & Q-DiT$^\dagger$ {\scriptsize\citep{qdit}} & 3.753 & 0.7235 & 3.566 & 1.438 & 0.7865 & 2.869 & 8.701 & 0.7182 & 2.607 \\
 & QuaRot$^\dagger$ {\scriptsize\citep{quarot}} & 2.917 & 0.7168 & 3.488 & 1.353 & 0.7835 & 2.764 & 7.192 & 0.7615 & 2.536 \\
 & GPTQ$^\dagger$ {\scriptsize\citep{gptq}} & 3.769 & 0.7235 & 3.543 & 1.601 & 0.7881 & 2.885 & 9.432 & 0.7227 & 2.636 \\
 & AWQ$^\dagger$ {\scriptsize\citep{awq}} & 3.495 & 0.7315 & 3.625 & 1.410 & 0.7948 & 2.953 & 8.109 & 0.7765 & 2.736 \\
 & SmoothQuant$^\dagger$ {\scriptsize\citep{smoothquant}} & 21.577 & 0.4985 & 2.367 & 8.336 & 0.6405 & 2.456 & 15.545 & 0.6124 & 2.567 \\
\cmidrule(lr){2-11}
 & TTS-Q & \textbf{2.475} & \textbf{0.7475} & \textbf{3.730} & \textbf{1.244} & \textbf{0.8023} & \textbf{2.990} & \textbf{7.158} & \textbf{0.7826} & \textbf{2.794} \\
\midrule
LongCat-3.5B & Full Prec & 1.952 & 0.7848 & 3.758 & 1.136 & 0.8195 & 3.189 & 6.702 & 0.7966 & 3.021 \\
\cmidrule(lr){2-11}
 & RTN & 113.786 & 0.0430 & 1.332 & 100.682 & -0.0097 & 1.367 & 100.194 & -0.0150 & 1.399 \\
 & Q-DiT$^\dagger$ {\scriptsize\citep{qdit}} & 2.138 & 0.7631 & 3.616 & 1.199 & 0.8048 & 2.996 & 7.105 & 0.7827 & 2.772 \\
 & QuaRot$^\dagger$ {\scriptsize\citep{quarot}} & 2.233 & 0.7587 & 3.542 & 1.215 & 0.7990 & 2.859 & 6.407 & 0.7807 & 2.632 \\
 & GPTQ$^\dagger$ {\scriptsize\citep{gptq}} & 2.238 & 0.7634 & 3.600 & 1.220 & 0.8047 & 2.989 & 7.071 & 0.7805 & 2.769 \\
 & AWQ$^\dagger$ {\scriptsize\citep{awq}} & \textbf{1.973} & 0.7689 & 3.628 & 1.227 & 0.8093 & 3.021 & 6.393 & 0.7898 & 2.815 \\
 & SmoothQuant$^\dagger$ {\scriptsize\citep{smoothquant}} & 4.782 & 0.5843 & 2.802 & 1.855 & 0.7375 & 2.504 & 7.606 & 0.7011 & 2.345 \\
\cmidrule(lr){2-11}
 & TTS-Q & 2.067 & \textbf{0.7730} & \textbf{3.683} & \textbf{1.140} & \textbf{0.8112} & \textbf{3.022} & \textbf{5.984} & \textbf{0.7925} & \textbf{2.833} \\
\midrule
F5-TTS & Full Prec & 1.932 & 0.6686 & 3.692 & 1.756 & 0.7596 & 2.944 & 7.391 & 0.7368 & 2.691 \\
\cmidrule(lr){2-11}
 & RTN & 109.229 & 0.1435 & 1.282 & 96.901 & 0.1044 & 1.318 & 97.805 & 0.0870 & 1.314 \\
 & Q-DiT$^\dagger$ {\scriptsize\citep{qdit}} & 2.414 & 0.5988 & 2.347 & 1.934 & 0.6981 & 1.756 & 7.492 & 0.6682 & 1.517 \\
 & QuaRot$^\dagger$ {\scriptsize\citep{quarot}} & 2.104 & 0.6140 & 2.207 & 1.835 & 0.7035 & 1.603 & 7.714 & 0.6762 & 1.433 \\
 & GPTQ$^\dagger$ {\scriptsize\citep{gptq}} & 2.351 & 0.6118 & 2.303 & 1.897 & 0.7056 & 1.746 & 7.775 & 0.6793 & 1.531 \\
 & AWQ$^\dagger$ {\scriptsize\citep{awq}} & 2.373 & 0.6012 & 2.274 & 1.903 & 0.7005 & 1.775 & 7.556 & 0.6759 & 1.557 \\
 & SmoothQuant$^\dagger$ {\scriptsize\citep{smoothquant}} & 16.232 & 0.4743 & 1.945 & 3.855 & 0.6725 & 2.103 & 10.306 & 0.6511 & 1.870 \\
\cmidrule(lr){2-11}
 & TTS-Q & \textbf{2.045} & \textbf{0.6541} & \textbf{3.457} & \textbf{1.758} & \textbf{0.7470} & \textbf{2.777} & \textbf{7.468} & \textbf{0.7233} & \textbf{2.523} \\
\bottomrule
\end{tabular}
\caption{Speech quality on Seed-TTS under W4A4 quantization of DiT linear layers. WER/CER are percentages. Full Prec uses each model's native precision; bold marks the best quantized value before rounding. Calibration-based methods share the selected conditions. $\dagger$ denotes a task adaptation (Appendix~\ref{sec:adapters}). Reconstruction controls are reported in Tables~\ref{tab:selection} and~\ref{tab:components}.}
\label{tab:main}
\end{table*}

\subsection{Experimental Setup}
\paragraph{Models and data.}
We evaluate LongCat-AudioDiT-1B and -3.5B \citep{longcat}, which generate waveform latents, and F5-TTS \citep{f5tts}, which generates mel-spectrograms. Each retains its native sampling and decoding procedure. The Seed-TTS evaluation set \citep{seed_tts,seed_eval} provides 1,088 English, 2,020 Mandarin, and 400 hard Mandarin conditions. Calibration uses 64 conditions selected from a common pool of 256 candidates drawn from separate development sources. A separate development set of 100 conditions per language is used to choose the coverage objective, learning rates, identity weight, and final checkpoint. These choices are fixed before test evaluation; test recordings are not used for optimization. Appendix~\ref{sec:implementation} details the data partitions.

\paragraph{Calibration budget.}
Each selected condition contributes eight time positions and two guidance branches, giving 512 condition--time states and 1,024 branch records. Each reconstruction unit receives one pass over these records in the first stage and one additional pass in the second. TTS-Q evaluates the identity objective once per condition during the second pass, for 64 evaluations.

\paragraph{Evaluation.}
We measure intelligibility with English WER and Mandarin CER, speaker similarity with SIM, and predicted naturalness with UTMOS \citep{utmos}. All metrics are utterance-level macro averages; methods share evaluation conditions and generation noise. SIM measures similarity to the reference prompt using the Seed toolkit's WavLM-based verifier \citep{wavlm}. The calibration loss instead compares student and teacher speech using ECAPA-TDNN \citep{ecapa}. Paired intervals for the loss ablations resample voice-prompt recordings, as detailed in Appendix~\ref{sec:implementation}.

\paragraph{Comparisons.}
We compare with RTN and adaptations of Q-DiT, QuaRot, GPTQ, AWQ, and SmoothQuant. All methods use W4A4 for the same DiT linear layers, with other operations in native precision. GPTQ, AWQ, and SmoothQuant include added dynamic four-bit activation quantization. Methods requiring calibration share each model's selected conditions but follow their own calibration procedures; Appendix~\ref{sec:adapters} specifies the adaptations. The ablations use our FlatQuant-style backend \citep{flatquant}, holding the quantization mechanism fixed while varying selection and supervision. Full-test quality uses simulated quantization; storage is measured separately from packed checkpoints.

\subsection{Quantitative Comparison}
TTS-Q consistently achieves higher SIM and UTMOS than the evaluated external baselines (Table~\ref{tab:main}), while generally reducing text error. AWQ on LongCat-3.5B English is the exception for text accuracy. Text accuracy alone can obscure differences in speech quality. On F5 English, QuaRot reaches a WER of 2.104, close to TTS-Q's 2.045, yet its UTMOS is 2.207 compared with 3.457. TTS-Q also remains below full precision in SIM and UTMOS on every subset, even where its text error is lower.

Much of the advantage over external baselines is already present with random-set reconstruction using our FlatQuant-style backend. On F5 English, it reaches 3.433 UTMOS, compared with 2.347 for the strongest external baseline on this metric. We therefore assess selection and identity supervision with the backend held fixed in Table~\ref{tab:components}.

Relative to reconstruction with randomly selected conditions, the complete procedure improves SIM and UTMOS across all evaluated model--subset pairs. Text error is lower in most settings, with F5 Mandarin as the exception and 1B hard Mandarin unchanged at the reported precision. We next examine the contributions of condition selection and identity supervision.

\begin{table}[!tb]
\centering\footnotesize
\setlength{\tabcolsep}{0.8pt}
\begin{tabular*}{\columnwidth}{@{\extracolsep{\fill}}llcccc@{}}
\toprule
&&EN&\multicolumn{3}{c}{ZH}\\
\cmidrule(lr){3-3}\cmidrule(lr){4-6}
Model & Selection & WER$\downarrow$ & CER$\downarrow$ & SIM$\uparrow$ & UTMOS$\uparrow$\\
\midrule
1B & Random$^{\S}$ & 2.592 & 1.271 & 0.8002 & 2.909 \\
 & Farthest-point$^*$ & \textbf{2.344} & \textbf{1.186} & \textbf{0.8006} & \textbf{2.953} \\
 & Average-distance & 2.598 & 1.262 & 0.7996 & 2.931 \\
\midrule
3.5B & Random$^{\S}$ & 2.089 & 1.152 & 0.8088 & 2.995 \\
 & Farthest-point & 2.140 & 1.214 & 0.8097 & 2.988 \\
 & Average-distance$^*$ & \textbf{1.979} & \textbf{1.137} & \textbf{0.8105} & \textbf{3.008} \\
\midrule
F5 & Random$^{\S}$ & 2.137 & 1.724 & 0.7461 & 2.721 \\
 & Farthest-point & 2.200 & 1.754 & \textbf{0.7472} & \textbf{2.774} \\
 & Average-distance$^*$ & \textbf{2.101} & \textbf{1.701} & 0.7470 & 2.746 \\
\bottomrule
\end{tabular*}
\caption{Calibration-set selection under the same reconstruction procedure and budget. $\S$~Mean of three random calibration sets. $^*$~Strategy selected on the development set for TTS-Q. Bold marks the best value per model before rounding. Full results appear in Appendix~\ref{sec:extended-results}.}
\label{tab:selection}
\end{table}

\begin{table*}[!t]
\centering\footnotesize
\setlength{\tabcolsep}{1.5pt}
\renewcommand{\arraystretch}{1.12}
\begin{tabular*}{\textwidth}{@{\extracolsep{\fill}}ll*{9}{r}@{}}
\toprule
&&\multicolumn{3}{c}{Seed English}&\multicolumn{3}{c}{Seed Mandarin}&\multicolumn{3}{c}{Seed Mandarin hard}\\
\cmidrule(lr){3-5}\cmidrule(lr){6-8}\cmidrule(lr){9-11}
Model & Variant & WER$\downarrow$ & SIM$\uparrow$ & UTMOS$\uparrow$ & CER$\downarrow$ & SIM$\uparrow$ & UTMOS$\uparrow$ & CER$\downarrow$ & SIM$\uparrow$ & UTMOS$\uparrow$\\
\midrule
1B & Random$^{\S}$ & 2.592 & 0.7466 & 3.673 & 1.271 & 0.8002 & 2.909 & 7.158 & 0.7813 & 2.727 \\
 & Selected coverage & \textbf{2.344} & 0.7435 & 3.674 & \textbf{1.186} & 0.8006 & 2.953 & \textbf{7.018} & 0.7816 & 2.757 \\
 & Reconstruction continuation & 2.518 & 0.7468 & 3.679 & 1.299 & 0.8001 & 2.940 & 7.516 & 0.7814 & 2.737 \\
 & TTS-Q & 2.475 & \textbf{0.7475} & \textbf{3.730} & 1.244 & \textbf{0.8023} & \textbf{2.990} & 7.158 & \textbf{0.7826} & \textbf{2.794} \\
\midrule
3.5B & Random$^{\S}$ & 2.089 & 0.7715 & 3.652 & 1.152 & 0.8088 & 2.995 & 6.382 & 0.7906 & 2.814 \\
 & Selected coverage & \textbf{1.979} & 0.7725 & 3.651 & \textbf{1.137} & 0.8105 & 3.008 & 6.387 & 0.7913 & 2.821 \\
 & Reconstruction continuation & 2.068 & 0.7723 & 3.641 & 1.162 & 0.8099 & 3.000 & 6.531 & 0.7911 & 2.802 \\
 & TTS-Q & 2.067 & \textbf{0.7730} & \textbf{3.683} & 1.140 & \textbf{0.8112} & \textbf{3.022} & \textbf{5.984} & \textbf{0.7925} & \textbf{2.833} \\
\midrule
F5 & Random$^{\S}$ & 2.137 & 0.6540 & 3.433 & 1.724 & 0.7461 & 2.721 & 7.752 & 0.7228 & 2.453 \\
 & Selected coverage & 2.101 & 0.6531 & 3.466 & \textbf{1.701} & \textbf{0.7470} & 2.746 & \textbf{7.370} & \textbf{0.7252} & 2.475 \\
 & Reconstruction continuation & 2.051 & 0.6519 & \textbf{3.526} & 1.713 & 0.7427 & 2.761 & 7.574 & 0.7220 & 2.480 \\
 & TTS-Q & \textbf{2.045} & \textbf{0.6541} & 3.457 & 1.758 & 0.7470 & \textbf{2.777} & 7.468 & 0.7233 & \textbf{2.523} \\
\bottomrule
\end{tabular*}
\caption{Complete component comparison on Seed-TTS. WER/CER are percentages. Selected coverage is the first-stage checkpoint, using farthest-point selection for 1B and average-distance selection for 3.5B and F5. TTS-Q applies the second stage with identity supervision; Reconstruction continuation ablates this supervision ($\lambda=0$). Both branches start from the same checkpoint with identical learning rates and update schedules. $\S$~Mean of three random calibration sets. Bold marks the best value per model and metric before rounding.}
\label{tab:components}
\end{table*}

\subsection{Calibration Set Selection}
\label{sec:selection-analysis}
We compare selection rules under the same reconstruction procedure, calibration size, and time-state budget (Table~\ref{tab:selection}). The coverage objective is chosen on the development set before test evaluation: farthest-point for 1B and average-distance for 3.5B and F5 (Appendix Table~\ref{tab:development-selection}). Random averages three independently calibrated subsets, with individual results reported in Appendix Table~\ref{tab:random-draws}.

\begin{figure}[!htb]
\centering
\input{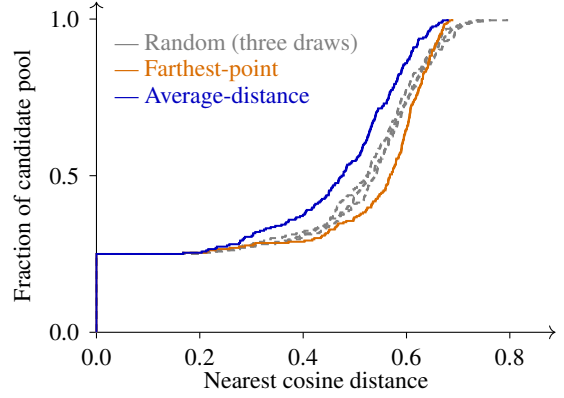}
\caption{Coverage of the calibration pool in the original speaker embedding space. Cumulative distributions of cosine distance to the nearest selected condition, using 64 of 256 candidates. Curves farther left indicate closer coverage. Gray dashed curves show the three random draws; the initial 25\% at zero comprises selected conditions.}
\label{fig:coverage-distance}
\end{figure}

Selection changes downstream quality even with the reconstruction procedure held fixed. The clearest gains in text accuracy occur at 1B. Farthest-point selection lowers English WER to 2.344, compared with 2.568--2.613 across random draws, and Mandarin CER to 1.186, compared with 1.252--1.295. Its English SIM, however, falls below all three draws. For 3.5B, the selected average-distance configuration improves English WER over every random draw; for F5, its clearest gain in text accuracy is on hard Mandarin.

Lower geometric coverage costs do not always yield better speech quality. Figure~\ref{fig:coverage-distance} shows that farthest-point selection reduces the largest coverage gaps while increasing the average distance relative to random selection. Average-distance selection reduces both costs on this pool (Appendix Table~\ref{tab:coverage-geometry-stats}), yet farthest-point gives better text accuracy for 1B. On F5, average-distance gives lower text error than farthest-point across the test subsets, at the cost of some Mandarin SIM and UTMOS. We therefore use speech quality on the development set to choose between the coverage objectives. Full test metrics appear in Appendix~\ref{sec:extended-results}.

\subsection{Speaker Identity Distillation}
\label{sec:identity-analysis}
The second stage applies identity supervision through further calibration. To separate its effect from that of additional reconstruction updates, we ablate the identity term by setting $\lambda=0$ (Reconstruction continuation in Table~\ref{tab:components}). Both branches start from the same Selected coverage checkpoint and use identical data, learning rates, and update schedules. This control isolates the loss term under the chosen schedule; it does not establish the best independently tuned reconstruction procedure. Comparison with the initial checkpoint measures the net change from the second stage. Appendix Table~\ref{tab:intervals} provides paired intervals for both comparisons.

Adding identity supervision raises SIM point estimates across all settings relative to continuation with reconstruction alone, though intervals exclude zero only on standard Mandarin. These gains are measured against the reference prompt with WavLM, independently of the ECAPA teacher--student objective used for calibration. UTMOS improves in all but F5 English, with intervals excluding zero for each improvement. Text accuracy changes less consistently; only the CER reduction on 3.5B hard Mandarin has an interval excluding zero.

\begin{table*}[t]
\begin{minipage}[t]{\columnwidth}
\vspace{0pt}
\centering
\footnotesize
\setlength{\tabcolsep}{2.5pt}
\renewcommand{\arraystretch}{1.20}
\begin{tabular}{llrrr}
\toprule
\multirow{2}{*}{Model} & \multirow{2}{*}{Scope} & \multicolumn{2}{c}{File size (MiB)} & \multirow{2}{*}{\shortstack{Compression\\ratio}} \\
\cmidrule(lr){3-4}
 & & FP16$^*$ & TTS-Q & \\
\midrule
1B & DiT Linear & 1871.1 & 476.1 & 3.93$\times$ \\
 & Pipeline & 3268.6 & 1873.6 & 1.74$\times$ \\
\midrule
3.5B & DiT Linear & 6473.5 & 1635.9 & 3.96$\times$ \\
 & Pipeline & 7875.2 & 3037.7 & 2.59$\times$ \\
\midrule
F5 & DiT Linear & 632.1 & 161.6 & 3.91$\times$ \\
 & Pipeline & 695.0 & 224.4 & 3.10$\times$ \\
\bottomrule
\end{tabular}
\caption{Measured file sizes; ratios divide reference by INT4-packed size. DiT Linear includes auxiliary parameters; Pipeline includes all synthesis components. $^*$The reference uses FP16 for in-scope weights; other tensors retain identical native precision in both bundles.}
\label{tab:storage}
\end{minipage}\hfill
\begin{minipage}[t]{\columnwidth}
\vspace{0pt}
\centering\footnotesize
\begin{tikzpicture}[x=0.97cm,y=0.88cm,font=\scriptsize]
\definecolor{deploygreen}{RGB}{98,133,80}
\definecolor{fpgray}{RGB}{192,194,196}
\draw[fill=fpgray,draw=black!60] (1.10,3.88) rectangle +(0.20,0.14);
\node[anchor=west] at (1.35,3.95) {Full Prec};
\draw[fill=deploygreen,draw=black!60] (3.65,3.88) rectangle +(0.20,0.14);
\node[anchor=west] at (3.90,3.95) {W4A4 deployment};
\begin{scope}[xshift=0cm]
\node at (1.45,3.52) {(a) Mean RTF};
\draw[black!65] (0,0) -- (2.95,0);
\draw[black!65] (0,0) -- (0,3.10);
\draw[black!12] (0,0.0000) -- (2.95,0.0000);
\node[anchor=east] at (-0.06,0.0000) {0.0};
\draw[black!12] (0,0.9688) -- (2.95,0.9688);
\node[anchor=east] at (-0.06,0.9688) {0.1};
\draw[black!12] (0,1.9375) -- (2.95,1.9375);
\node[anchor=east] at (-0.06,1.9375) {0.2};
\draw[black!12] (0,2.9062) -- (2.95,2.9062);
\node[anchor=east] at (-0.06,2.9062) {0.3};
\draw[fill=fpgray,draw=black!60,line width=.25pt] (0.1550,0) rectangle (0.4450,1.2840);
\node[anchor=south,font=\fontsize{6}{7}\selectfont,inner sep=1pt] at (0.3000,1.3090) {0.133};
\draw[fill=deploygreen,draw=black!60,line width=.25pt] (0.5150,0) rectangle (0.8050,0.7747);
\node[anchor=south,font=\fontsize{6}{7}\selectfont,inner sep=1pt] at (0.6600,0.7997) {0.080};
\node[anchor=north] at (0.4800,-.06) {1B};
\draw[fill=fpgray,draw=black!60,line width=.25pt] (1.1450,0) rectangle (1.4350,2.7139);
\node[anchor=south,font=\fontsize{6}{7}\selectfont,inner sep=1pt] at (1.2900,2.7389) {0.280};
\draw[fill=deploygreen,draw=black!60,line width=.25pt] (1.5050,0) rectangle (1.7950,1.1587);
\node[anchor=south,font=\fontsize{6}{7}\selectfont,inner sep=1pt] at (1.6500,1.1837) {0.120};
\node[anchor=north] at (1.4700,-.06) {3.5B};
\draw[fill=fpgray,draw=black!60,line width=.25pt] (2.1350,0) rectangle (2.4250,0.6454);
\node[anchor=south,font=\fontsize{6}{7}\selectfont,inner sep=1pt] at (2.2800,0.6704) {0.067};
\draw[fill=deploygreen,draw=black!60,line width=.25pt] (2.4950,0) rectangle (2.7850,1.4015);
\node[anchor=south,font=\fontsize{6}{7}\selectfont,inner sep=1pt] at (2.6400,1.4265) {0.145};
\node[anchor=north] at (2.4600,-.06) {F5};
\node[rotate=90] at (-0.65,1.55) {RTF $\downarrow$};
\end{scope}
\begin{scope}[xshift=3.95cm]
\node at (1.45,3.52) {(b) Peak allocated memory};
\draw[black!65] (0,0) -- (2.95,0);
\draw[black!65] (0,0) -- (0,3.10);
\draw[black!12] (0,0.0000) -- (2.95,0.0000);
\node[anchor=east] at (-0.06,0.0000) {0};
\draw[black!12] (0,0.9118) -- (2.95,0.9118);
\node[anchor=east] at (-0.06,0.9118) {5};
\draw[black!12] (0,1.8235) -- (2.95,1.8235);
\node[anchor=east] at (-0.06,1.8235) {10};
\draw[black!12] (0,2.7353) -- (2.95,2.7353);
\node[anchor=east] at (-0.06,2.7353) {15};
\draw[fill=fpgray,draw=black!60,line width=.25pt] (0.1550,0) rectangle (0.4450,1.0536);
\node[anchor=south,font=\fontsize{6}{7}\selectfont,inner sep=1pt] at (0.3000,1.0786) {5.78};
\draw[fill=deploygreen,draw=black!60,line width=.25pt] (0.5150,0) rectangle (0.8050,0.5556);
\node[anchor=south,font=\fontsize{6}{7}\selectfont,inner sep=1pt] at (0.6600,0.5806) {3.05};
\node[anchor=north] at (0.4800,-.06) {1B};
\draw[fill=fpgray,draw=black!60,line width=.25pt] (1.1450,0) rectangle (1.4350,2.7461);
\node[anchor=south,font=\fontsize{6}{7}\selectfont,inner sep=1pt] at (1.2900,2.7711) {15.06};
\draw[fill=deploygreen,draw=black!60,line width=.25pt] (1.5050,0) rectangle (1.7950,0.9739);
\node[anchor=south,font=\fontsize{6}{7}\selectfont,inner sep=1pt] at (1.6500,0.9989) {5.34};
\node[anchor=north] at (1.4700,-.06) {3.5B};
\draw[fill=fpgray,draw=black!60,line width=.25pt] (2.1350,0) rectangle (2.4250,0.1515);
\node[anchor=south,font=\fontsize{6}{7}\selectfont,inner sep=1pt] at (2.2800,0.1765) {0.83};
\draw[fill=deploygreen,draw=black!60,line width=.25pt] (2.4950,0) rectangle (2.7850,0.1729);
\node[anchor=south,font=\fontsize{6}{7}\selectfont,inner sep=1pt] at (2.6400,0.3529) {0.95};
\node[anchor=north] at (2.4600,-.06) {F5};
\node[rotate=90] at (-0.65,1.55) {GiB $\downarrow$};
\end{scope}
\end{tikzpicture}

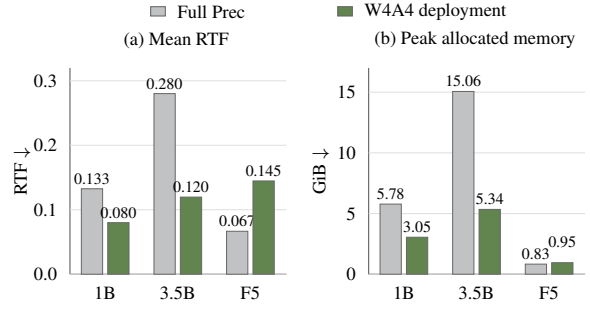
\captionof{figure}{Warm H100 inference on EN30+ZH30, repeated three times with CUDA Graphs on both paths. Full Prec uses FP32 with TF32 disabled (LongCat) or FP16 (F5). Peak allocated memory includes model and graph allocations. Execution details and exact values appear in Appendix~\ref{sec:deployment}.}
\label{fig:runtime}
\end{minipage}
\end{table*}

Relative to the starting checkpoint, calibration with identity supervision raises both SIM and UTMOS point estimates on every LongCat subset. On F5 Mandarin, its role is to retain the similarity achieved in the first stage: continuation with reconstruction alone lowers SIM from 0.7470 to 0.7427, while TTS-Q maintains 0.7470 at the reported precision. The two quality metrics need not move together: on F5 English, identity supervision raises SIM relative to reconstruction-only continuation but lowers UTMOS by 0.069.

Some text accuracy gained in the first stage is lost during the second: WER/CER point estimates rise in most settings relative to the starting checkpoint, although the intervals for these increases include zero. F5 hard Mandarin additionally loses some SIM while gaining UTMOS.

\subsection{Storage and Runtime}
\label{sec:storage}
INT4 packing reduces complete-pipeline storage by 1.74--3.10$\times$ relative to the FP16 weight-storage reference (Table~\ref{tab:storage}). Unquantized components limit savings compared with the 3.91--3.96$\times$ compression of DiT linear layers. The teacher and speaker encoder add no inference storage. Appendix~\ref{sec:storage-accounting} details the serialized contents.

LongCat executes four-bit codes in INT8 containers; F5 uses floating-point W4A4 fusion. With CUDA Graphs on both paths, LongCat achieves warm H100 speedups of 1.66$\times$ and 2.34$\times$ over native FP32, reducing peak allocated memory by 47.3\% and 64.5\% (Figure~\ref{fig:runtime}). F5 remains slower and uses more allocated memory than native FP16.

On the shared EN30+ZH30 subset, LongCat deployment changes SIM and UTMOS only slightly, with mixed text-error changes; F5 outputs match the original W4A4 path sample for sample. This audit does not reproduce full-test quality with integer execution (Appendix~\ref{sec:deployment}).

\section{Conclusion}
TTS-Q combines calibration conditions selected for speaker coverage with identity supervision from a full-precision teacher. At W4A4, it achieves higher speaker similarity and predicted naturalness than the evaluated external baselines. It also improves both metrics over reconstruction with randomly selected conditions, while generally reducing text error.

The choice of coverage objective depends on the model. Under matched continuation, identity supervision raises speaker similarity point estimates across all settings. Relative to the first-stage checkpoint, the second stage improves both similarity and predicted naturalness on LongCat but gives mixed results on F5, and some initial gains in text accuracy are lost.

INT4 packing reduces storage for the complete pipeline by 1.74--3.10$\times$ relative to the FP16 weight-storage reference. Integer execution also accelerates warm LongCat inference by 1.66--2.34$\times$ over its native FP32 reference and reduces allocated memory. F5 retains a floating-point implementation with runtime overhead.

\section*{Limitations}
Listening tests are needed to confirm the benefits indicated by automatic metrics. Evaluation covers two model families in English and Mandarin at W4A4. Selection uses one candidate pool and budget; generalization to other pools, budgets, and speaker representations remains untested.

Speech quality is evaluated with one generation seed. Random selection averages three sets, but selected coverage and its continuations use one fixed set. The intervals do not account for variation across calibration runs or configuration selection. Prompt-level resampling may also underestimate uncertainty when recordings share a speaker (Appendix~\ref{sec:implementation}).

Identity gradients cover only the final two sampling updates; F5 also substitutes a cached teacher prefix for the student's trajectory. The effects of these approximations on final quality have not been isolated.

Deployment measurements cover warm H100 inference on 60 conditions. LongCat uses INT8 execution of four-bit codes, not native packed INT4 kernels. Full-test quality with integer execution, cold-start acceleration, and on-device performance remain unverified.
\section*{Ethical Considerations}
Zero-shot speech synthesis can reproduce a voice from a short recording. Reducing model storage requirements can broaden access to speech synthesis, including its misuse for impersonation. This study concerns quantization of existing models and does not establish consent to clone any speaker. Any released audio should respect the source datasets' licenses and permissions and be identified as synthetic. Voice similarity measures model fidelity, not identity authentication.
\clearpage
\bibliography{references}

@inproceedings{f5tts,
  title={{F5-TTS: A Fairytaler that Fakes Fluent and Faithful Speech with Flow Matching}},
  author={Chen, Yushen and Niu, Zhikang and Ma, Ziyang and Deng, Keqi and Wang, Chunhui and Zhao, Jian and Yu, Kai and Chen, Xie},
  booktitle={Proceedings of the 63rd Annual Meeting of the Association for Computational Linguistics (Volume 1: Long Papers)},
  pages={6255--6271},
  year={2025},
  doi={10.18653/v1/2025.acl-long.313},
  url={https://aclanthology.org/2025.acl-long.313/}
}

@article{longcat,
  title={{LongCat-AudioDiT: High-Fidelity Diffusion Text-to-Speech in the Waveform Latent Space}},
  author={Xin, Detai and Hu, Shujie and Yang, Chengzuo and Huang, Chen and Yu, Guoqiao and Wan, Guanglu and Cai, Xunliang},
  year={2026},
  journal={arXiv preprint arXiv:2603.29339},
  url={https://arxiv.org/abs/2603.29339}
}

@inproceedings{gptq,
  title={{GPTQ: Accurate Post-Training Quantization for Generative Pre-trained Transformers}},
  author={Frantar, Elias and Ashkboos, Saleh and Hoefler, Torsten and Alistarh, Dan},
  booktitle={The Eleventh International Conference on Learning Representations},
  year={2023},
  url={https://openreview.net/forum?id=tcbBPnfwxS}
}

@inproceedings{awq,
  title={{AWQ: Activation-aware Weight Quantization for On-Device LLM Compression and Acceleration}},
  author={Lin, Ji and Tang, Jiaming and Tang, Haotian and Yang, Shang and Chen, Wei-Ming and Wang, Wei-Chen and Xiao, Guangxuan and Dang, Xingyu and Gan, Chuang and Han, Song},
  booktitle={Proceedings of Machine Learning and Systems},
  volume={6},
  year={2024},
  url={https://proceedings.mlsys.org/paper_files/paper/2024/hash/42a452cbafa9dd64e9ba4aa95cc1ef21-Abstract-Conference.html}
}

@inproceedings{smoothquant,
  title={{SmoothQuant: Accurate and Efficient Post-Training Quantization for Large Language Models}},
  author={Xiao, Guangxuan and Lin, Ji and Seznec, Mickael and Wu, Hao and Demouth, Julien and Han, Song},
  booktitle={Proceedings of the 40th International Conference on Machine Learning},
  series={Proceedings of Machine Learning Research},
  volume={202},
  pages={38087--38099},
  year={2023},
  url={https://proceedings.mlr.press/v202/xiao23c.html}
}

@inproceedings{quarot,
  title={{QuaRot: Outlier-Free 4-Bit Inference in Rotated LLMs}},
  author={Ashkboos, Saleh and Mohtashami, Amirkeivan and Croci, Maximilian L. and Li, Bo and Cameron, Pashmina and Jaggi, Martin and Alistarh, Dan and Hoefler, Torsten and Hensman, James},
  booktitle={Advances in Neural Information Processing Systems},
  volume={37},
  year={2024},
  doi={10.52202/079017-3180},
  url={https://proceedings.neurips.cc/paper_files/paper/2024/hash/b5b939436789f76f08b9d0da5e81af7c-Abstract-Conference.html}
}

@inproceedings{flatquant,
  title={{FlatQuant: Flatness Matters for LLM Quantization}},
  author={Sun, Yuxuan and Liu, Ruikang and Bai, Haoli and Bao, Han and Zhao, Kang and Li, Yuening and Hu, Jiaxin and Yu, Xianzhi and Hou, Lu and Yuan, Chun and Jiang, Xin and Liu, Wulong and Yao, Jun},
  booktitle={Proceedings of the 42nd International Conference on Machine Learning},
  series={Proceedings of Machine Learning Research},
  volume={267},
  pages={57587--57613},
  year={2025},
  url={https://proceedings.mlr.press/v267/sun25l.html}
}

@inproceedings{ptq4dit,
  title={{PTQ4DiT: Post-training Quantization for Diffusion Transformers}},
  author={Wu, Junyi and Wang, Haoxuan and Shang, Yuzhang and Shah, Mubarak and Yan, Yan},
  booktitle={Advances in Neural Information Processing Systems},
  volume={37},
  year={2024},
  url={https://proceedings.neurips.cc/paper_files/paper/2024/hash/72d32f4fe0b7af03732bd227bf1c4a5f-Abstract-Conference.html}
}

@inproceedings{qdit,
  title={{Q-DiT: Accurate Post-Training Quantization for Diffusion Transformers}},
  author={Chen, Lei and Meng, Yuan and Tang, Chen and Ma, Xinzhu and Jiang, Jingyan and Wang, Xin and Wang, Zhi and Zhu, Wenwu},
  booktitle={Proceedings of the IEEE/CVF Conference on Computer Vision and Pattern Recognition},
  pages={28306--28315},
  year={2025},
  url={https://openaccess.thecvf.com/content/CVPR2025/html/Chen_Q-DiT_Accurate_Post-Training_Quantization_for_Diffusion_Transformers_CVPR_2025_paper.html}
}

@inproceedings{audio_ptq,
  title={{Post-Training Quantization for Audio Diffusion Transformers}},
  author={Khandelwal, Tanmay and Fuentes, Magdalena},
  booktitle={2025 IEEE Workshop on Applications of Signal Processing to Audio and Acoustics},
  pages={1--5},
  year={2025},
  doi={10.1109/WASPAA66052.2025.11230962},
  url={https://doi.org/10.1109/WASPAA66052.2025.11230962}
}

@inproceedings{ptq4adm,
  title={{PTQ4ADM: Post-Training Quantization for Efficient Text Conditional Audio Diffusion Models}},
  author={Vora, Jayneel and Krishnan, Aditya and Bouacida, Nader and Shankar, Prabhu R. V. and Mohapatra, Prasant},
  booktitle={2025 IEEE International Conference on Acoustics, Speech and Signal Processing},
  pages={1--5},
  year={2025},
  doi={10.1109/ICASSP49660.2025.10890341},
  url={https://ieeexplore.ieee.org/document/10890341/}
}

@inproceedings{viditq,
  title={{ViDiT-Q: Efficient and Accurate Quantization of Diffusion Transformers for Image and Video Generation}},
  author={Zhao, Tianchen and Fang, Tongcheng and Huang, Haofeng and Wan, Rui and Soedarmadji, Widyadewi and Liu, Enshu and Li, Shiyao and Lin, Zinan and Dai, Guohao and Yan, Shengen and Yang, Huazhong and Ning, Xuefei and Wang, Yu},
  booktitle={International Conference on Learning Representations},
  year={2025},
  url={https://openreview.net/forum?id=E1N1oxd63b}
}

@inproceedings{qvdit,
  title={{Q-VDiT: Towards Accurate Quantization and Distillation of Video-Generation Diffusion Transformers}},
  author={Feng, Weilun and Yang, Chuanguang and Qin, Haotong and Li, Xiangqi and Wang, Yu and An, Zhulin and Huang, Libo and Diao, Boyu and Zhao, Zixiang and Xu, Yongjun and Magno, Michele},
  booktitle={Proceedings of the 42nd International Conference on Machine Learning},
  series={Proceedings of Machine Learning Research},
  volume={267},
  pages={16956--16976},
  year={2025},
  url={https://proceedings.mlr.press/v267/feng25q.html}
}

@inproceedings{s2qvdit,
  title={{S$^2$Q-VDiT: Accurate Quantized Video Diffusion Transformer with Salient Data and Sparse Token Distillation}},
  author={Feng, Weilun and Qin, Haotong and Yang, Chuanguang and Li, Xiangqi and Yang, Han and Li, Yuqi and An, Zhulin and Huang, Libo and Magno, Michele and Xu, Yongjun},
  booktitle={Advances in Neural Information Processing Systems},
  volume={38},
  year={2025},
  doi={10.52202/085713-1172},
  url={https://papers.neurips.cc/paper_files/paper/2025/hash/31ed129feae64a7e44a15b148c15558d-Abstract-Conference.html}
}

@inproceedings{bittts,
  title={{BitTTS: Highly Compact Text-to-Speech Using 1.58-bit Quantization and Weight Indexing}},
  author={Kawamura, Masaya and Hasumi, Takuya and Shirahata, Yuma and Yamamoto, Ryuichi},
  booktitle={Interspeech 2025},
  pages={5538--5542},
  year={2025},
  doi={10.21437/Interspeech.2025-1122},
  url={https://www.isca-archive.org/interspeech_2025/kawamura25_interspeech.html}
}

@inproceedings{self_calibration,
  title={{Self-calibration for Language Model Quantization and Pruning}},
  author={Williams, Miles and Chrysostomou, George and Aletras, Nikolaos},
  booktitle={Proceedings of the 2025 Conference of the Nations of the Americas Chapter of the Association for Computational Linguistics: Human Language Technologies (Volume 1: Long Papers)},
  pages={10149--10167},
  year={2025},
  doi={10.18653/v1/2025.naacl-long.509},
  url={https://aclanthology.org/2025.naacl-long.509/}
}

@inproceedings{ecapa,
  title={{ECAPA-TDNN: Emphasized Channel Attention, Propagation and Aggregation in TDNN Based Speaker Verification}},
  author={Desplanques, Brecht and Thienpondt, Jenthe and Demuynck, Kris},
  booktitle={Interspeech 2020},
  pages={3830--3834},
  year={2020},
  doi={10.21437/Interspeech.2020-2650},
  url={https://www.isca-archive.org/interspeech_2020/desplanques20_interspeech.html}
}

@article{utmos,
  title={{UTMOS: UTokyo-SaruLab System for VoiceMOS Challenge 2022}},
  author={Saeki, Takaaki and Xin, Detai and Nakata, Wataru and Koriyama, Tomoki and Takamichi, Shinnosuke and Saruwatari, Hiroshi},
  year={2022},
  journal={arXiv preprint arXiv:2204.02152},
  url={https://arxiv.org/abs/2204.02152}
}

@misc{seed_eval,
 author={{ByteDance Speech Team}},
 title={{seed-tts-eval: Objective Evaluation Toolkit and Test Set for Seed-TTS}},
 year={2024},
 url={https://github.com/BytedanceSpeech/seed-tts-eval}
}

@article{voicebox,
  title={{Voicebox: Text-Guided Multilingual Universal Speech Generation at Scale}},
  author={Le, Matthew and Vyas, Apoorv and Shi, Bowen and Karrer, Brian and Sari, Leda and Moritz, Rashel and Williamson, Mary and Manohar, Vimal and Adi, Yossi and Mahadeokar, Jay and Hsu, Wei-Ning},
  year={2023},
  journal={arXiv preprint arXiv:2306.15687},
  url={https://arxiv.org/abs/2306.15687}
}

@article{e2tts,
  title={{E2 TTS: Embarrassingly Easy Fully Non-Autoregressive Zero-Shot TTS}},
  author={Eskimez, Sefik Emre and Wang, Xiaofei and Thakker, Manthan and Li, Canrun and Tsai, Chung-Hsien and Xiao, Zhen and Yang, Hemin and Zhu, Zirun and Tang, Min and Tan, Xu and Liu, Yanqing and Zhao, Sheng and Kanda, Naoyuki},
  year={2024},
  journal={arXiv preprint arXiv:2406.18009},
  url={https://arxiv.org/abs/2406.18009}
}

@article{valle,
  title={{Neural Codec Language Models are Zero-Shot Text to Speech Synthesizers}},
  author={Wang, Chengyi and Chen, Sanyuan and Wu, Yu and Zhang, Ziqiang and Zhou, Long and Liu, Shujie and Chen, Zhuo and Liu, Yanqing and Wang, Huaming and Li, Jinyu and He, Lei and Zhao, Sheng and Wei, Furu},
  year={2023},
  journal={arXiv preprint arXiv:2301.02111},
  url={https://arxiv.org/abs/2301.02111}
}

@article{naturalspeech2,
  title={{NaturalSpeech 2: Latent Diffusion Models are Natural and Zero-Shot Speech and Singing Synthesizers}},
  author={Shen, Kai and Ju, Zeqian and Tan, Xu and Liu, Yanqing and Leng, Yichong and He, Lei and Qin, Tao and Zhao, Sheng and Bian, Jiang},
  year={2023},
  journal={arXiv preprint arXiv:2304.09116},
  url={https://arxiv.org/abs/2304.09116}
}

@article{adaround,
  title={{Up or Down? Adaptive Rounding for Post-Training Quantization}},
  author={Nagel, Markus and Amjad, Rana Ali and van Baalen, Mart and Louizos, Christos and Blankevoort, Tijmen},
  year={2020},
  journal={arXiv preprint arXiv:2004.10568},
  url={https://arxiv.org/abs/2004.10568}
}

@article{brecq,
  title={{BRECQ: Pushing the Limit of Post-Training Quantization by Block Reconstruction}},
  author={Li, Yuhang and Gong, Ruihao and Tan, Xu and Yang, Yang and Hu, Peng and Zhang, Qi and Yu, Fengwei and Wang, Wei and Gu, Shi},
  year={2021},
  journal={arXiv preprint arXiv:2102.05426},
  url={https://arxiv.org/abs/2102.05426}
}

@inproceedings{qdiffusion,
  title={{Q-Diffusion: Quantizing Diffusion Models}},
  author={Li, Xiuyu and Liu, Yijiang and Lian, Long and Yang, Huanrui and Dong, Zhen and Kang, Daniel and Zhang, Shanghang and Keutzer, Kurt},
  booktitle={Proceedings of the IEEE/CVF International Conference on Computer Vision},
  pages={17535--17545},
  year={2023},
  url={https://openaccess.thecvf.com/content/ICCV2023/html/Li_Q-Diffusion_Quantizing_Diffusion_Models_ICCV_2023_paper.html}
}

@inproceedings{calibration_data,
  title={{On the Impact of Calibration Data in Post-training Quantization and Pruning}},
  author={Williams, Miles and Aletras, Nikolaos},
  booktitle={Proceedings of the 62nd Annual Meeting of the Association for Computational Linguistics (Volume 1: Long Papers)},
  year={2024},
  doi={10.18653/v1/2024.acl-long.544},
  url={https://aclanthology.org/2024.acl-long.544/}
}

@inproceedings{coreset,
  title={{Active Learning for Convolutional Neural Networks: A Core-Set Approach}},
  author={Sener, Ozan and Savarese, Silvio},
  booktitle={International Conference on Learning Representations},
  year={2018},
  url={https://openreview.net/forum?id=H1aIuk-RW}
}

@inproceedings{yourtts,
  title={{YourTTS: Towards Zero-Shot Multi-Speaker TTS and Zero-Shot Voice Conversion for Everyone}},
  author={Casanova, Edresson and Weber, Julian and Shulby, Christopher D and Junior, Arnaldo Candido and G{\"o}lge, Eren and Ponti, Moacir A},
  booktitle={Proceedings of the 39th International Conference on Machine Learning},
  series={Proceedings of Machine Learning Research},
  volume={162},
  pages={2709--2720},
  year={2022},
  url={https://proceedings.mlr.press/v162/casanova22a.html}
}

@inproceedings{draft,
  title={Directly Fine-Tuning Diffusion Models on Differentiable Rewards},
  author={Clark, Kevin and Vicol, Paul and Swersky, Kevin and Fleet, David J.},
  booktitle={International Conference on Learning Representations},
  year={2024},
  url={https://arxiv.org/abs/2309.17400}
}

@inproceedings{libritts_r,
  title={{LibriTTS-R: A Restored Multi-Speaker Text-to-Speech Corpus}},
  author={Koizumi, Yuma and Zen, Heiga and Karita, Shigeki and Ding, Yifan and Yatabe, Kohei and Morioka, Nobuyuki and Bacchiani, Michiel and Zhang, Yu and Han, Wei and Bapna, Ankur},
  booktitle={Interspeech 2023},
  pages={5496--5500},
  year={2023},
  doi={10.21437/Interspeech.2023-1584},
  url={https://www.isca-archive.org/interspeech_2023/koizumi23_interspeech.html}
}

@inproceedings{aishell1,
  title={{AISHELL-1: An Open-Source Mandarin Speech Corpus and a Speech Recognition Baseline}},
  author={Bu, Hui and Du, Jiayu and Na, Xingyu and Wu, Bengu and Zheng, Hao},
  booktitle={Oriental COCOSDA 2017},
  year={2017},
  url={https://arxiv.org/abs/1709.05522}
}

@inproceedings{aishell3,
  title={{AISHELL-3: A Multi-Speaker Mandarin TTS Corpus}},
  author={Shi, Yao and Bu, Hui and Xu, Xin and Zhang, Shaoji and Li, Ming},
  booktitle={Interspeech 2021},
  year={2021},
  url={https://www.isca-archive.org/interspeech_2021/shi21c_interspeech.html}
}

@misc{primewords,
  title={{Primewords Chinese Corpus Set 1}},
  author={{Primewords Information Technology Co., Ltd.}},
  year={2018},
  url={https://www.openslr.org/47/}
}

@inproceedings{whisper,
  title={{Robust Speech Recognition via Large-Scale Weak Supervision}},
  author={Radford, Alec and Kim, Jong Wook and Xu, Tao and Brockman, Greg and Mcleavey, Christine and Sutskever, Ilya},
  booktitle={Proceedings of the 40th International Conference on Machine Learning},
  series={Proceedings of Machine Learning Research},
  volume={202},
  pages={28492--28518},
  year={2023},
  url={https://proceedings.mlr.press/v202/radford23a.html}
}

@misc{whisper_large_v3,
  title={{Whisper large-v3 Model Card}},
  author={{OpenAI}},
  year={2023},
  url={https://huggingface.co/openai/whisper-large-v3}
}

@inproceedings{paraformer,
  title={{Paraformer: Fast and Accurate Parallel Transformer for Non-autoregressive End-to-End Speech Recognition}},
  author={Gao, Zhifu and Zhang, Shiliang and McLoughlin, Ian and Yan, Zhijie},
  booktitle={Interspeech 2022},
  pages={2063--2067},
  year={2022},
  doi={10.21437/Interspeech.2022-9996},
  url={https://www.isca-archive.org/interspeech_2022/gao22b_interspeech.html}
}

@article{wavlm,
  title={{WavLM: Large-Scale Self-Supervised Pre-Training for Full Stack Speech Processing}},
  author={Chen, Sanyuan and Wang, Chengyi and Chen, Zhengyang and Wu, Yu and Liu, Shujie and Chen, Zhuo and Li, Jinyu and Kanda, Naoyuki and Yoshioka, Takuya and Xiao, Xiong and Wu, Jian and Zhou, Long and Ren, Shuo and Qian, Yanmin and Qian, Yao and Wu, Jian and Zeng, Michael and Yu, Xiangzhan and Wei, Furu},
  journal={IEEE Journal of Selected Topics in Signal Processing},
  year={2022},
  doi={10.1109/JSTSP.2022.3188113},
  url={https://arxiv.org/abs/2110.13900}
}

@article{seed_tts,
  title={{Seed-TTS: A Family of High-Quality Versatile Speech Generation Models}},
  author={Anastassiou, Philip and Chen, Jiawei and Chen, Jitong and Chen, Yuanzhe and Chen, Zhuo and Chen, Ziyi and Cong, Jian and Deng, Lelai and Ding, Chuang and Gao, Lu and Gong, Mingqing and Huang, Peisong and Huang, Qingqing and Huang, Zhiying and Huo, Yuanyuan and Jia, Dongya and Li, Chumin and Li, Feiya and Li, Hui and Li, Jiaxin and Li, Xiaoyang and Li, Xingxing and Liu, Lin and Liu, Shouda and Liu, Sichao and Liu, Xudong and Liu, Yuchen and Liu, Zhengxi and Lu, Lu and Pan, Junjie and Wang, Xin and Wang, Yuping and Wang, Yuxuan and Wei, Zhen and Wu, Jian and Yao, Chao and Yang, Yifeng and Yi, Yuanhao and Zhang, Junteng and Zhang, Qidi and Zhang, Shuo and Zhang, Wenjie and Zhang, Yang and Zhao, Zilin and Zhong, Dejian and Zhuang, Xiaobin},
  journal={arXiv preprint arXiv:2406.02430},
  year={2024},
  url={https://arxiv.org/abs/2406.02430}
}
\clearpage
\appendix
\section{Implementation Details}
\label{sec:implementation}
\subsection{Model and Sampling Configurations}
LongCat-AudioDiT-1B and -3.5B use 15 Euler updates with adaptive projection guidance at scale 4.0. F5-TTS Base uses 32 Euler updates, classifier-free guidance at scale 2.0, and sway sampling with coefficient $-1$. We retain the official text processing, duration handling, and waveform decoder for each model.

W4A4 covers all 261, 341, and 167 DiT linear layers, respectively, including those outside the transformer blocks. Biases and all other operations retain their native precision. Full-test quality is evaluated using simulated quantization.

\subsection{Data Construction and Selection Protocol}
Development conditions are drawn from LibriTTS-R \citep{libritts_r}, AISHELL-1 \citep{aishell1}, AISHELL-3 \citep{aishell3}, and Primewords \citep{primewords}. The development pool contains 1,000 English and 1,000 Mandarin conditions, and a separate development set contains 100 conditions per language. The development set is used to specify the calibration-set selection strategy and select the learning rates, identity weight $\lambda$, and final checkpoint. All choices are frozen before final evaluation on Seed-TTS. These partitions are separated by speaker and source-associated groups. Source families used for testing are excluded from the development data.

We form a fixed candidate pool of 256 conditions, taking 32 from each of two languages and four duration strata, and select 64 conditions from it. Each condition contributes eight sampled time positions and two guidance branches: 512 condition--time states and 1,024 branch records. All models share the candidate conditions and their reference embeddings; only the coverage objective differs.

Evaluation uses the full Seed-TTS subsets listed in Section~\ref{sec:experiments} and the official evaluation toolkit \citep{seed_eval}. Test recordings are excluded from calibration inputs.

\subsection{Evaluation and Statistical Reporting}
Evaluation uses Whisper-large-v3 \citep{whisper,whisper_large_v3} for English WER, Paraformer \citep{paraformer} for Mandarin CER, WavLM \citep{wavlm} for speaker similarity, and UTMOS22 \citep{utmos} for predicted naturalness. SIM compares generated speech with the reference prompt. The calibration loss instead compares student and teacher speech using ECAPA-TDNN, so evaluation does not reuse the representation optimized by the loss. WER/CER are percentages, and UTMOS is an automatic estimate rather than a listening-test score.

Metrics are macro-averaged over utterances. Methods share evaluation conditions and generation noise derived from base seed 42. Failed generations and scoring failures remain in the run accounting; incomplete evaluations are not reported as complete means. All nine RTN evaluations completed without either type of failure. Their error rates can exceed 100\% because insertion errors are not capped.

For the continuation comparisons, we pair outputs by condition and bootstrap voice-prompt recordings with replacement 10{,}000 times. All utterances sharing a prompt stay together in each resample. The 2.5th and 97.5th percentiles give the intervals in Table~\ref{tab:intervals}. The manifests lack audited speaker identities, so different recordings of one speaker may be resampled independently. These are prompt-cluster intervals for fixed configurations; they do not account for configuration selection or variation across calibration runs. An interval containing zero does not establish the absence of an effect.

Random denotes the unweighted mean of three independently calibrated random subsets, computed from unrounded utterance-level means. Table~\ref{tab:random-draws} reports each draw and the sample standard deviation across draws. Selected coverage and its continuations use one fixed set. Variation across random draws measures sensitivity to the calibration set, whereas the paired intervals measure uncertainty in comparisons using the same set.

\subsection{Optimization Settings}
Initial reconstruction uses AdamW with zero weight decay, a cosine schedule, and gradient-norm clipping at 1. The initial learning rates are $5\times10^{-3}$ for transforms and channel scales and $5\times10^{-2}$ for clipping parameters. Each reconstruction unit receives one pass over the 1,024 branch records. The final checkpoint is the one selected on the development set before Seed-TTS evaluation.

The second calibration stage starts from the selected-coverage checkpoint and uses the same 64 conditions and time-state records for one additional pass. In TTS-Q, the speaker objective is evaluated 64 times, once per condition, through the final two updates of each generated segment. The reconstruction-only ablation independently starts from the same checkpoint and sets the identity weight to zero, retaining identical learning rates and update schedules. 

LongCat-1B uses farthest-point coverage, a continuation transform/scale learning rate of $2.5\times10^{-4}$, clipping learning rate of $2.5\times10^{-3}$, and identity weight $\lambda=0.2$. LongCat-3.5B uses average-distance coverage, learning rates $5\times10^{-4}$ and $5\times10^{-3}$, and identity weight $\lambda=0.1$. F5 uses average-distance coverage, transform/scale and clipping learning rates of $5\times10^{-4}$ and $5\times10^{-3}$, and identity weight $\lambda=0.2$.

\subsection{Reconstruction Backend}
Our reconstruction backend follows the learned-transform construction of FlatQuant \citep{flatquant}. For a row-vector input $x$ and a weight matrix $W$, an invertible input transform $T$ preserves the floating-point linear map:
\begin{equation}
 xW^\top=(xT)(WT^{-\top})^\top.
\end{equation}
We quantize the transformed activation and weight on the symmetric integer grid $[-7,7]$, using a dynamic per-token activation scale and a per-output-channel weight scale. Each linear layer has its own Kronecker-factored transform, channel scales, and clipping parameters.

Local reconstruction follows Equation~\ref{eq:rec-detail} in Section~\ref{sec:coverage}, using cached teacher inputs for every reconstruction unit. The same targets are reused during the second calibration stage.

\subsection{Backpropagation through Sampling}
Identity features use frozen ECAPA-TDNN after differentiable resampling to 16\,kHz. For $N$ sampling updates, waveform decoder $V$, and quantized transitions $\Psi_{\phi,n}$ incorporating the time schedule and guidance, the calibration path is
\begin{equation}
 \begin{aligned}
 \widetilde z_{N-H}&=\operatorname{sg}[z^{\mathrm{src}}_{N-H}],\\
 \widetilde z_{n+1}&=\Psi_{\phi,n}(\widetilde z_n,c),\\
 \widetilde G_\phi(c)&=V(\widetilde z_N),
 \end{aligned}
 \label{eq:suffix}
\end{equation}
for $n=N-H,\ldots,N-1$, with $H=2$. LongCat obtains $z^{\mathrm{src}}_{N-H}$ from the current student trajectory without recording gradients, then reruns the final two updates with gradients. For F5, the equation applies to each segment with a cached teacher starting state; the decoded segments are crossfaded before computing the identity loss. Detaching the starting state limits backpropagation to the final two updates in both models, whereas teacher-state reuse also removes the student prefix computation in F5.

All three configurations shuffle branch-state visits across conditions. For each condition, the identity gradient is injected at the unconditional-branch visit of its highest-indexed selected sampling position, which need not be the final sampling step. F5 indexes these positions across segment trajectories. This choice specifies the injection time only; the identity loss uses both guidance branches in the final two updates.

At this visit, a single backward evaluation with all quantized modules enabled computes identity gradients at the current parameters, before any unit updates. These gradients are detached and held fixed while each unit combines its corresponding slice with its local reconstruction gradient:
\begin{equation}
 g_b=\nabla_{\phi_b}\ell_b+
       \lambda\nabla_{\phi_b}\mathcal L_{\mathrm{id}}(c;\phi).
 \label{eq:gradient}
\end{equation}
Here the identity term is evaluated at the parameters from the start of the visit. The combined gradient is clipped per unit before its optimizer step, and the identity gradients are discarded after this visit. Other visits use only reconstruction gradients. Quantization uses a straight-through estimator, and the pretrained DiT, decoder, and speaker encoder remain fixed.

\subsection{Numerical Validation}
Validation checks cover full-precision state replay, equivalence with quantization disabled, finite gradients, checkpoint reloads, quantization coverage, and unchanged pretrained weights. Differentiable speaker-feature centering is checked against the original implementation. For numerical stability, the identity loss is scaled by 256 during backpropagation and its gradients unscaled before combination, leaving $\lambda$ unchanged. Configurations, generated audio, per-utterance scores, and asset hashes are retained.

\def\baselineappendixblock{%
\begin{minipage}{\textwidth}
\section{Baseline Adaptations}
\label{sec:adapters}
Table~\ref{tab:baseline-settings} summarizes the external baselines in Table~\ref{tab:main} and the FlatQuant-style reconstruction control in Tables~\ref{tab:selection} and~\ref{tab:components}. GPTQ, AWQ, and SmoothQuant share each model's selected 64 calibration conditions. The reconstruction ablations use 64 conditions under the selection rule specified in each row. RTN, Q-DiT, and the local QuaRot adapter require no offline calibration. All methods use the linear scope defined in Appendix~\ref{sec:implementation}.

\centering\footnotesize
\renewcommand{\arraystretch}{1.12}
\setlength{\tabcolsep}{3pt}
\begin{tabularx}{\linewidth}{@{}l X@{}}
\toprule
Method & Quantization configuration\\
\midrule
RTN & Symmetric W4 per output channel; dynamic A4 per token; codes in $[-7,7]$.\\
\addlinespace
Q-DiT & Official dynamic group quantization with width 128, without granularity search.\\
\addlinespace
QuaRot & Local randomized Hadamard transform with round-to-nearest quantization; no residual-stream rotation or GPTQ.\\
\addlinespace
GPTQ & Official Hessian compensation; block/group width 128; damping 0.01; no activation ordering; added dynamic group-128 A4.\\
\addlinespace
AWQ & Official 20-point scale search and weight clipping; explicit inverse scaling; no query/key clipping; added dynamic group-128 A4.\\
\addlinespace
SmoothQuant & Official scale migration with $\alpha=0.5$; explicit inverse scaling in place of normalization fusion; symmetric per-output-channel W4; added dynamic group-128 A4.\\
\addlinespace
FlatQuant (ablation) & Local Kronecker-transform adapter; reconstruction on teacher trajectory states with 64 conditions chosen by the stated selection rule. This backend is shared by initial reconstruction and both continuation branches in Table~\ref{tab:components}, with the same W4A4 quantizer as TTS-Q.\\
\bottomrule
\end{tabularx}
\captionof{table}{Baseline and reconstruction-control configurations; Q-DiT, QuaRot, GPTQ, AWQ, SmoothQuant, and FlatQuant are task-adapted as specified.}
\label{tab:baseline-settings}
\end{minipage}
}
\twocolumn[\baselineappendixblock\vspace{0.8\baselineskip}]

Group quantizers allow a smaller final group when a layer width is not divisible by 128. The Hadamard adapter pads non-power-of-two widths to the next power of two.

\par\smallskip
\noindent\begin{minipage}{\columnwidth}
\section{Coverage Geometry Diagnostics}
\label{sec:coverage-geometry-diagnostics}

We assign each pool candidate to its nearest selected condition. If $n_j$ candidates are assigned to condition $j$, cell balance is measured by the total-variation distance $\frac{1}{2}\sum_{j\in S}|n_j/|\mathcal P|-1/K|$. Max cell is $\max_{j\in S}n_j$. These statistics describe the geometry of the selected set.

\centering\footnotesize
\setlength{\tabcolsep}{2.5pt}
\begin{tabular}{lrrrr}
\toprule
Selection & $C_{\max}$ & $C_{\mathrm{avg}}$ & Max cell & TV \\
\midrule
Random draw 0    & 0.795 & 0.412 & 13 & 0.258 \\
Random draw 1    & 0.761 & 0.413 & 12 & 0.258 \\
Random draw 2    & 0.739 & 0.417 & 15 & 0.242 \\
Farthest-point   & 0.689 & 0.429 & 12 & 0.230 \\
Average-distance & 0.680 & 0.376 & 10 & 0.195 \\
\bottomrule
\end{tabular}
\captionof{table}{Geometry of the 64-of-256 selections in reference-speaker embedding space. Max cell is the largest nearest-assignment cell; TV is the distance from cell masses to uniform mass.}
\label{tab:coverage-geometry-stats}
\end{minipage}

\section{Storage Measurement}
\label{sec:storage-accounting}
Table~\ref{tab:storage} reports actual file sizes at two scopes: quantized DiT linear layers and the complete synthesis pipeline. One MiB is $2^{20}$ bytes. The complete bundle includes unquantized DiT state, LongCat's UMT5 and VAE or F5's Vocos, persistent buffers, tokenizer assets, and model configurations. Calibration-only teachers and speaker encoders are excluded.

For each quantized layer, the learned balancing and inverse transform are folded into the weights before packing two signed INT4 codes per byte. The bundle also stores FP32 weight scales, channel-balancing values, forward Kronecker factors, and activation-clipping parameters. Weight clipping is absorbed into the scales; no floating-point copy of the packed weights is retained.

The reference stores in-scope linear weights in FP16. All remaining tensors retain their native precision and are identical in both bundles. This reference is used only for storage accounting: the Full Prec quality results use native FP32 DiT weights for LongCat and FP16 for F5. Shared tensors are stored once.

Sizes include safetensors headers and padding, configuration files, and the state index, without entropy compression. We verify complete coverage of model state and exact CPU round trips for all 769 quantized layers and all unquantized tensors. Storage sizes do not measure runtime memory; Figure~\ref{fig:runtime} reports inference speed separately.

\clearpage
\onecolumn
\section{Extended Results}
\label{sec:extended-results}

Table~\ref{tab:extended-selection} reports the full selection results on the three Seed subsets; the complete component comparison appears in Table~\ref{tab:components} in the main text. Table~\ref{tab:intervals} gives paired differences against the selected-coverage checkpoint and the matched reconstruction continuation. In Table~\ref{tab:extended-selection}, bold marks the best unrounded value per model and metric; WER/CER are percentages.

\begingroup
\centering\footnotesize
\setlength{\tabcolsep}{4pt}
\renewcommand{\arraystretch}{1.12}
\begin{tabular*}{\textwidth}{@{\extracolsep{\fill}}ll*{9}{r}@{}}
\toprule
&&\multicolumn{3}{c}{Seed English}&\multicolumn{3}{c}{Seed Mandarin}&\multicolumn{3}{c}{Seed Mandarin hard}\\
\cmidrule(lr){3-5}\cmidrule(lr){6-8}\cmidrule(lr){9-11}
Model & Variant & WER$\downarrow$ & SIM$\uparrow$ & UTMOS$\uparrow$ & CER$\downarrow$ & SIM$\uparrow$ & UTMOS$\uparrow$ & CER$\downarrow$ & SIM$\uparrow$ & UTMOS$\uparrow$\\
\midrule
1B & Random$^{\S}$ & 2.592 & \textbf{0.7466} & 3.673 & 1.271 & 0.8002 & 2.909 & 7.158 & 0.7813 & 2.727 \\
 & Farthest-point & \textbf{2.344} & 0.7435 & \textbf{3.674} & \textbf{1.186} & \textbf{0.8006} & \textbf{2.953} & \textbf{7.018} & \textbf{0.7816} & \textbf{2.757} \\
 & Average-distance & 2.598 & 0.7448 & 3.674 & 1.262 & 0.7996 & 2.931 & 7.386 & 0.7805 & 2.741 \\
\midrule
3.5B & Random$^{\S}$ & 2.089 & 0.7715 & \textbf{3.652} & 1.152 & 0.8088 & 2.995 & \textbf{6.382} & 0.7906 & 2.814 \\
 & Farthest-point & 2.140 & \textbf{0.7726} & 3.651 & 1.214 & 0.8097 & 2.988 & 6.941 & 0.7902 & 2.787 \\
 & Average-distance & \textbf{1.979} & 0.7725 & 3.651 & \textbf{1.137} & \textbf{0.8105} & \textbf{3.008} & 6.387 & \textbf{0.7913} & \textbf{2.821} \\
\midrule
F5 & Random$^{\S}$ & 2.137 & \textbf{0.6540} & 3.433 & 1.724 & 0.7461 & 2.721 & 7.752 & 0.7228 & 2.453 \\
 & Farthest-point & 2.200 & 0.6537 & 3.418 & 1.754 & \textbf{0.7472} & \textbf{2.774} & 7.469 & 0.7245 & \textbf{2.509} \\
 & Average-distance & \textbf{2.101} & 0.6531 & \textbf{3.466} & \textbf{1.701} & 0.7470 & 2.746 & \textbf{7.370} & \textbf{0.7252} & 2.475 \\
\bottomrule
\end{tabular*}
\captionof{table}{Full metrics for random selection and the two coverage objectives. Random is the mean of three draws; all rows use reconstruction without speaker identity distillation.}
\label{tab:extended-selection}
\par\endgroup

\par\vspace{0.8\baselineskip}
\normalsize\raggedright
\subsection{Paired Comparisons}
\begingroup
\centering\footnotesize
\setlength{\tabcolsep}{0.75pt}
\renewcommand{\arraystretch}{1.02}
\begin{tabular*}{\textwidth}{@{\extracolsep{\fill}}lllrrr@{}}
\toprule
Model & Reference & Subset & $\Delta$ WER/CER (pp) & $\Delta$ SIM & $\Delta$ UTMOS\\
\midrule
1B & \shortstack[l]{Reconstruction\\continuation} & English & $-0.043$ $[-0.293,+0.219]$ & $+0.0007$ $[-0.0010,+0.0024]$ & $+0.052$ $[+0.042,+0.061]^{*}$ \\
 &  & Mandarin & $-0.055$ $[-0.185,+0.071]$ & $+0.0022$ $[+0.0013,+0.0031]^{*}$ & $+0.050$ $[+0.044,+0.057]^{*}$ \\
 &  & Mandarin hard & $-0.358$ $[-1.361,+0.280]$ & $+0.0011$ $[-0.0007,+0.0030]$ & $+0.058$ $[+0.044,+0.072]^{*}$ \\
\midrule
1B & \shortstack[l]{Selected\\coverage} & English & $+0.131$ $[-0.201,+0.455]$ & $+0.0040$ $[+0.0022,+0.0059]^{*}$ & $+0.056$ $[+0.047,+0.066]^{*}$ \\
 &  & Mandarin & $+0.058$ $[-0.070,+0.184]$ & $+0.0017$ $[+0.0007,+0.0026]^{*}$ & $+0.038$ $[+0.031,+0.044]^{*}$ \\
 &  & Mandarin hard & $+0.140$ $[-0.238,+0.511]$ & $+0.0010$ $[-0.0008,+0.0028]$ & $+0.037$ $[+0.023,+0.051]^{*}$ \\
\midrule
3.5B & \shortstack[l]{Reconstruction\\continuation} & English & $-0.001$ $[-0.192,+0.219]$ & $+0.0007$ $[-0.0007,+0.0021]$ & $+0.042$ $[+0.032,+0.052]^{*}$ \\
 &  & Mandarin & $-0.022$ $[-0.146,+0.097]$ & $+0.0013$ $[+0.0005,+0.0021]^{*}$ & $+0.022$ $[+0.016,+0.028]^{*}$ \\
 &  & Mandarin hard & $-0.546$ $[-1.242,-0.050]^{*}$ & $+0.0014$ $[-0.0003,+0.0032]$ & $+0.032$ $[+0.018,+0.045]^{*}$ \\
\midrule
3.5B & \shortstack[l]{Selected\\coverage} & English & $+0.087$ $[-0.123,+0.330]$ & $+0.0005$ $[-0.0009,+0.0020]$ & $+0.032$ $[+0.021,+0.043]^{*}$ \\
 &  & Mandarin & $+0.003$ $[-0.101,+0.106]$ & $+0.0007$ $[-0.0002,+0.0016]$ & $+0.013$ $[+0.007,+0.019]^{*}$ \\
 &  & Mandarin hard & $-0.403$ $[-0.675,-0.135]^{*}$ & $+0.0013$ $[-0.0006,+0.0032]$ & $+0.013$ $[-0.001,+0.026]$ \\
\midrule
F5 & \shortstack[l]{Reconstruction\\continuation} & English & $-0.006$ $[-0.190,+0.184]$ & $+0.0022$ $[-0.0000,+0.0045]$ & $-0.069$ $[-0.084,-0.054]^{*}$ \\
 &  & Mandarin & $+0.045$ $[-0.057,+0.147]$ & $+0.0043$ $[+0.0031,+0.0055]^{*}$ & $+0.017$ $[+0.007,+0.027]^{*}$ \\
 &  & Mandarin hard & $-0.106$ $[-0.486,+0.254]$ & $+0.0013$ $[-0.0013,+0.0039]$ & $+0.042$ $[+0.025,+0.060]^{*}$ \\
\midrule
F5 & \shortstack[l]{Selected\\coverage} & English & $-0.056$ $[-0.253,+0.126]$ & $+0.0010$ $[-0.0014,+0.0033]$ & $-0.010$ $[-0.025,+0.006]$ \\
 &  & Mandarin & $+0.057$ $[-0.059,+0.171]$ & $-0.0001$ $[-0.0013,+0.0012]$ & $+0.031$ $[+0.022,+0.040]^{*}$ \\
 &  & Mandarin hard & $+0.098$ $[-0.280,+0.457]$ & $-0.0019$ $[-0.0044,+0.0006]$ & $+0.047$ $[+0.029,+0.066]^{*}$ \\
\bottomrule
\end{tabular*}
\captionof{table}{Paired prompt-cluster bootstrap for TTS-Q minus each reference on the full Seed subsets. Reconstruction continuation is the second-stage ablation with identity weight set to zero; Selected coverage is the common first-stage checkpoint from which both branches start. Entries give mean differences and 95\% intervals; $^{*}$ marks intervals excluding zero. Negative WER/CER differences and positive SIM/UTMOS differences favor TTS-Q. The resampling protocol is given in Appendix~\ref{sec:implementation}.}
\label{tab:intervals}
\par\endgroup

\clearpage
\subsection{Random Calibration Draws}
Table~\ref{tab:random-draws} reports three independently calibrated random subsets of 64 conditions with the synthesis seed fixed. Their mean supplies the Random reference throughout the paper; their range describes calibration-set variation rather than a confidence interval.

\par\medskip
\begingroup
\centering\footnotesize
\setlength{\tabcolsep}{4pt}
\renewcommand{\arraystretch}{1.12}
\begin{tabular*}{\textwidth}{@{\extracolsep{\fill}}ll*{9}{r}@{}}
\toprule
&&\multicolumn{3}{c}{Seed English}&\multicolumn{3}{c}{Seed Mandarin}&\multicolumn{3}{c}{Seed Mandarin hard}\\
\cmidrule(lr){3-5}\cmidrule(lr){6-8}\cmidrule(lr){9-11}
Model & Draw & WER$\downarrow$ & SIM$\uparrow$ & UTMOS$\uparrow$ & CER$\downarrow$ & SIM$\uparrow$ & UTMOS$\uparrow$ & CER$\downarrow$ & SIM$\uparrow$ & UTMOS$\uparrow$\\
\midrule
1B & 0 & 2.613 & 0.7480 & 3.644 & 1.252 & 0.8005 & 2.884 & 7.518 & 0.7817 & 2.699 \\
 & 1 & 2.596 & 0.7452 & 3.671 & 1.265 & 0.7999 & 2.920 & 6.864 & 0.7804 & 2.731 \\
 & 2 & 2.568 & 0.7465 & 3.702 & 1.295 & 0.8002 & 2.923 & 7.093 & 0.7817 & 2.753 \\
 & Mean & 2.592 & 0.7466 & 3.673 & 1.271 & 0.8002 & 2.909 & 7.158 & 0.7813 & 2.727 \\
 & SD & 0.022 & 0.0014 & 0.029 & 0.022 & 0.0003 & 0.022 & 0.332 & 0.0007 & 0.027 \\
\midrule
3.5B & 0 & 2.169 & 0.7720 & 3.634 & 1.102 & 0.8090 & 2.986 & 6.393 & 0.7901 & 2.806 \\
 & 1 & 2.110 & 0.7716 & 3.661 & 1.205 & 0.8087 & 3.004 & 6.308 & 0.7916 & 2.829 \\
 & 2 & 1.989 & 0.7709 & 3.661 & 1.147 & 0.8086 & 2.995 & 6.446 & 0.7901 & 2.807 \\
 & Mean & 2.089 & 0.7715 & 3.652 & 1.152 & 0.8088 & 2.995 & 6.382 & 0.7906 & 2.814 \\
 & SD & 0.092 & 0.0005 & 0.016 & 0.052 & 0.0002 & 0.009 & 0.070 & 0.0009 & 0.013 \\
\midrule
F5 & 0 & 2.136 & 0.6527 & 3.459 & 1.675 & 0.7448 & 2.752 & 7.832 & 0.7215 & 2.477 \\
 & 1 & 2.224 & 0.6573 & 3.382 & 1.776 & 0.7478 & 2.661 & 7.735 & 0.7248 & 2.401 \\
 & 2 & 2.050 & 0.6519 & 3.458 & 1.720 & 0.7455 & 2.750 & 7.689 & 0.7220 & 2.483 \\
 & Mean & 2.137 & 0.6540 & 3.433 & 1.724 & 0.7461 & 2.721 & 7.752 & 0.7228 & 2.453 \\
 & SD & 0.087 & 0.0029 & 0.044 & 0.050 & 0.0016 & 0.052 & 0.073 & 0.0018 & 0.046 \\
\bottomrule
\end{tabular*}
\captionof{table}{Individual random calibration draws, their means, and sample standard deviations (SD, denominator $3-1$). Mean rows supply Random throughout the paper. WER/CER are percentages; generation seed 42 is fixed.}
\label{tab:random-draws}
\par\endgroup

\par\medskip
\normalsize\raggedright
\section{Development-Set Coverage Selection}
\label{sec:development-selection}
Table~\ref{tab:development-selection} reports the development-set results used to choose each model's coverage objective before Seed-TTS evaluation. The development partition is described in Appendix~\ref{sec:implementation}. All variants use reconstruction-based calibration without speaker identity distillation. Random denotes the mean over three random calibration sets.

Farthest-point selection performs better than average-distance selection on all reported metrics for LongCat-1B. Average-distance selection performs better on all metrics for F5-TTS and all except English UTMOS for LongCat-3.5B. Accordingly, the selected configurations use farthest-point coverage for LongCat-1B and average-distance coverage for LongCat-3.5B and F5-TTS.

Relative to random selection, the chosen coverage configuration improves all reported development metrics for LongCat-3.5B, all except English UTMOS for LongCat-1B, and all except English SIM for F5-TTS.

\par\medskip
\begin{minipage}{\textwidth}
\centering\footnotesize
\setlength{\tabcolsep}{4pt}
\renewcommand{\arraystretch}{1.12}
\begin{tabular*}{\textwidth}{@{\extracolsep{\fill}}ll*{6}{r}@{}}
\toprule
&&\multicolumn{3}{c}{Development English}&\multicolumn{3}{c}{Development Mandarin}\\
\cmidrule(lr){3-5}\cmidrule(lr){6-8}
Model & Selection & WER$\downarrow$ & SIM$\uparrow$ & UTMOS$\uparrow$ & CER$\downarrow$ & SIM$\uparrow$ & UTMOS$\uparrow$\\
\midrule
LongCat-1B & Random & 3.563 & 0.7721 & \textbf{4.053} & 14.144 & 0.7634 & 2.516 \\
 & Farthest-point$^*$ & \textbf{3.522} & \textbf{0.7793} & 4.039 & \textbf{14.123} & \textbf{0.7635} & \textbf{2.541} \\
 & Average-distance & 3.737 & 0.7654 & 3.997 & 14.952 & 0.7586 & 2.530 \\
\midrule
LongCat-3.5B & Random & 4.093 & 0.7793 & 3.836 & 7.888 & 0.7721 & 2.531 \\
 & Farthest-point & 3.932 & 0.7792 & \textbf{3.884} & 8.434 & 0.7683 & 2.547 \\
 & Average-distance$^*$ & \textbf{3.870} & \textbf{0.7828} & 3.849 & \textbf{7.034} & \textbf{0.7730} & \textbf{2.555} \\
\midrule
F5-TTS & Random & 4.404 & \textbf{0.6890} & 4.096 & 9.265 & 0.6950 & 2.087 \\
 & Farthest-point & 4.466 & 0.6875 & 4.127 & 9.131 & 0.7006 & 2.094 \\
 & Average-distance$^*$ & \textbf{4.157} & 0.6876 & \textbf{4.152} & \textbf{9.085} & \textbf{0.7268} & \textbf{2.235} \\
\bottomrule
\end{tabular*}
\captionof{table}{Calibration-set selection on the development set. WER/CER are percentages; bold marks the best reported value per model and metric. $^*$ marks the coverage strategy chosen for subsequent calibration and test evaluation. Random is the mean of three draws.}
\label{tab:development-selection}
\end{minipage}

\clearpage
\section{Deployment Evaluation}
\label{sec:deployment}
\paragraph{Execution paths.}
LongCat retains the learned transforms, scales, clipping parameters, and four-bit codebook $[-7,7]$. Codes are stored in INT8 containers for integer matrix multiplication with INT32 accumulation, followed by output rescaling and bias addition. Kernels are tuned to matrix shapes, with caching and CUDA Graphs to reduce overhead. This is W4A4 quantization with INT8 execution, not native packed INT4 inference. F5 retains the original quantization and Kronecker-transform precision, using floating-point operator fusion and CUDA Graphs.

\paragraph{Measurement protocol.}
All models use the same 30 English and 30 Mandarin conditions, each repeated three times. Within each model and language, the native and deployment paths are tested sequentially on the same H100; both use CUDA Graphs. LongCat's native reference uses FP32 with TF32 disabled, and F5's uses FP16. Warm timing includes reference-audio reading and preprocessing, sampling, and decoding, excluding model loading, graph preparation, output writing, and scoring. RTF is latency divided by generated audio duration. Mean RTF weights every condition and repetition equally; speedup divides the native mean RTF by the deployment mean RTF.

Peak allocated memory is the maximum PyTorch allocation over warm calls. It includes resident model and graph allocations, excludes temporary peaks during separate graph preparation, and does not measure total GPU memory usage. Table~\ref{tab:deployment-timing} reports exact runtime measurements. First preparation takes 1.230/1.966 seconds for Full Prec/W4A4 on LongCat-1B and 2.351/2.581 seconds on LongCat-3.5B. These are not complete cold-start measurements; acceleration claims apply only to warm inference.

\begingroup
\centering\footnotesize
\setlength{\tabcolsep}{4pt}
\begin{tabular}{llrrrr}
\toprule
Model & Path & Mean latency (s) & Mean RTF & Speedup & Peak allocation (GiB) \\
\midrule
LongCat-1B & Full Prec & 0.6372 & 0.13254 & 1.00$\times$ & 5.778 \\
 & W4A4, INT8 & 0.3836 & 0.07997 & 1.66$\times$ & 3.047 \\
\midrule
LongCat-3.5B & Full Prec & 1.3728 & 0.28014 & 1.00$\times$ & 15.059 \\
 & W4A4, INT8 & 0.5772 & 0.11961 & 2.34$\times$ & 5.341 \\
\midrule
F5-TTS & Full Prec & 0.2880 & 0.06662 & 1.00$\times$ & 0.831 \\
 & W4A4, fused & 0.6237 & 0.14467 & 0.46$\times$ & 0.948 \\
\bottomrule
\end{tabular}
\captionof{table}{Exact deployment measurements underlying Figure~\ref{fig:runtime}. Speedup divides mean Full Prec RTF by mean W4A4 RTF.}
\label{tab:deployment-timing}
\par\endgroup

\paragraph{Quality audit.}
Table~\ref{tab:deployment-quality} compares LongCat's original W4A4 execution with integer execution on the same EN30+ZH30 subset. SIM and UTMOS change little, while text error changes in both directions. Paired 95\% intervals for SIM and UTMOS all include zero; this does not establish equivalence. Scored audio was verified against the LongCat deployment outputs. Separately, F5's fused and original W4A4 paths produce sample-identical audio on all 60 conditions. This audit does not reproduce the full-test quality evaluation with LongCat integer execution.

\begin{minipage}{\textwidth}
\centering\footnotesize
\setlength{\tabcolsep}{3pt}
\renewcommand{\arraystretch}{1.15}
\begin{tabular*}{\textwidth}{@{\extracolsep{\fill}}llrrrrrr@{}}
\toprule
& & \multicolumn{2}{c}{WER/CER (\%)} & \multicolumn{2}{c}{SIM} & \multicolumn{2}{c}{UTMOS} \\
\cmidrule(lr){3-4}\cmidrule(lr){5-6}\cmidrule(lr){7-8}
Model & Subset & Original $\to$ Integer & $\Delta$ (pp) & Original $\to$ Integer & $\Delta$ & Original $\to$ Integer & $\Delta$ \\
\midrule
1B & EN & $2.9933\to1.7855$ & $-1.2078$ & $0.7546\to0.7567$ & $+0.0021$ & $3.6838\to3.6774$ & $-0.0063$ \\
 & ZH & $1.3143\to1.5227$ & $+0.2083$ & $0.8123\to0.8149$ & $+0.0026$ & $3.0238\to3.0315$ & $+0.0077$ \\
\midrule
3.5B & EN & $1.4522\to1.6903$ & $+0.2381$ & $0.7767\to0.7749$ & $-0.0017$ & $3.7083\to3.6952$ & $-0.0131$ \\
 & ZH & $1.0179\to1.2140$ & $+0.1961$ & $0.8196\to0.8196$ & $-0.0001$ & $3.0043\to3.0082$ & $+0.0039$ \\
\bottomrule
\end{tabular*}
\captionof{table}{LongCat deployment quality on EN30+ZH30. Differences are integer execution minus original W4A4 execution, computed before rounding. These subset scores are separate from the full-test results in the main text.}
\label{tab:deployment-quality}
\end{minipage}

\end{document}